%% file: manuscript.tex
\documentclass[11pt]{article}

\usepackage{scicite}
\usepackage{tabularx}
\usepackage{array}
\usepackage{threeparttable}

\usepackage{url}
\usepackage{hyperref}
\usepackage{amsmath}
\usepackage{amsfonts}
\usepackage{amssymb}
\usepackage{changes} 
\usepackage{mathrsfs,adjustbox}
\usepackage{graphicx,subfigure}
\usepackage{wrapfig}
\usepackage{lscape}
\usepackage{rotating}
\usepackage{epstopdf}
\usepackage{longtable}
\usepackage{footnote}
\usepackage{tablefootnote}
\usepackage{booktabs,caption}
\usepackage{makecell} 
\usepackage{times}
\usepackage{ifthen}
\makeatletter
\renewcommand\@cite[2]{%
  [{#1\ifthenelse{\boolean{@tempswa}}{,\nolinebreak[3] #2}{}}]}
\makeatother

\providecommand{\keywords}[1]
{
  \small	
  \textbf{\textit{Keywords--}} #1
}

\usepackage[margin=2.0cm]{geometry}
\ifx
\fi

\newenvironment{sciabstract}{%
\begin{quote} \bf}
{\end{quote}}

\title{On The Importance of Fundamental Computational Fluid Dynamics Towards a Robust and Reliable Model of Left Atrial Flows: Is There More Than Meets the Eye?}

\author
{Ehsan Khalili,$^{1}$ C\'ecile Daversin-Catty,$^{2}$ Andy L. Olivares,$^{3}$ \\
Jordi Mill,$^{3}$ Oscar Camara,$^{3}$ Kristian Valen-Sendstad$^{1\ast}$
\\
\normalsize{$^{1}$Department of Computational Physiology,}\\
\normalsize{$^{}$Simula Research Laboratory, Oslo, Norway}\\
\normalsize{$^{2}$Department of Numerical Analysis and Scientific Computing, }\\
\normalsize{$^{}$Simula Research Laboratory, Oslo, Norway}\\
\normalsize{$^{3}$Department of Information and Communication Technologies,}\\
\normalsize{$^{}$Universitat Pompeu Fabra, Barcelona, Spain}\\
\\
\normalsize{$^\ast$ Corresponding author, E-mail: kvs@simula.no}
}
\date{}

\begin{document} 

\baselineskip12pt
\maketitle 

\begin{sciabstract}
\normalfont{Computational fluid dynamics (CFD) studies of left atrial flows have reached a sophisticated level, e.g., revealing plausible relationships between hemodynamics and stresses with atrial fibrillation. However, little focus has been on fundamental fluid modelling of LA flows. The purpose of this study was to investigate the spatiotemporal convergence, along with the differences between high- (HR) versus normal-resolution/accuracy (NR) solution strategies, respectively. CFD simulations on 12 patient-specific left atrial geometries, obtained from computed tomography scans, were performed by using a second-order accurate and space/time centered solver. The convergence studies showed an average variability of around 30\% and 55\% for time averaged wall shear stress (WSS), oscillatory shear index (OSI), relative residence time (RRT), and endothelial cell activation potential (ECAP), even between intermediate spatial and temporal resolutions, in the left atrium (LA) and left atrial appendage (LAA), respectively. The comparison between HR and NR simulations showed good correlation in the LA for WSS, RRT, and ECAP ($R^2>$ 0.9), but not for OSI ($R^2=$ 0.63). However, there were poor correlations in the LAA especially for OSI, RRT, and ECAP ($R^2=$ 0.55, 0.63, and 0.61, respectively), except for WSS ($R^2=$ 0.81). The errors are comparable to differences previously reported with disease correlations. To robustly predict atrial hemodynamics and stresses, numerical resolutions of 10M elements and 10k time-steps per cycle seem necessary (i.e., one order of magnitude higher than normally used in both space and time). In conclusion, attention to fundamental numerical aspects is essential towards establishing a plausible, robust, and reliable model of LA flows.}
\end{sciabstract}

\section{INTRODUCTION}\label{sec_intro}

Stroke is the leading cause of death worldwide, where cardiogenic emboli is the reason for 20\% to 40\% of all cases~\cite{hur2011dual}. Atrial fibrillation (AF), as the most common form of cardiac arrhythmia, disturbs electrical signals and subsequently prevents normal contractile function of the left atrium (LA), which affects dynamical behaviour of blood flow. Around 99\% of AF-related strokes originate from thrombus formed in the LA~\cite{cresti2019prevalence}, specifically in the left atrial appendage (LAA). Thrombus formation is normally explained by the \textit{Virchow’s triad}, i.e., hypercoagulability, presence of endothelial injury, and blood flow stasis.  

Intuitively, one would seek to image stagnant blood flow, but the velocity fields measured by imaging techniques such as echocardiography provide limited resolution, usually do not satisfy the governing equations of fluid dynamics, and suffer from partial volume effects~\cite{rispoli2015computational}, On the other hand, a high-resolution representation of cardiac anatomy is routinely available and the idea has been to numerically compute the blood flow instead~\cite{taylor2010image}. Computational fluid dynamics (CFD) has been used to retrospectively correlate blood flow in the LA with disease states in search of a prospective clinical tool to predict thrombus formation and supplement clinical risk scores like CHA$_2$DS$_2$-VASc~\cite{developed2010guidelines,sanatkhani2021subject}. The focus has been mostly on blood flow-induced stresses, which are believed to elicit distinct cellular responses linked to thrombus formation. The most commonly computed hemodynamic indices in the left atrial fluid modelling literature are time-averaged wall shear stress (WSS)~\cite{koizumi2015numerical, garcia2018sensitivity, zingaro2021hemodynamics, duenas2021comprehensive}, oscillatory shear index (OSI)~\cite{koizumi2015numerical, garcia2018sensitivity, paliwal2021presence, zingaro2021hemodynamics, duenas2021comprehensive}, relative residence time (RRT)~\cite{koizumi2015numerical, garcia2018sensitivity, paliwal2021presence, zingaro2021hemodynamics, duenas2021comprehensive}, and endothelial cell activation potential (ECAP)~\cite{garcia2018sensitivity, paliwal2021presence, duenas2021comprehensive, aguado2019silico, mill2021silico, corti2022impact}. Some have focused on correlating hemodynamics and different LAA morphologies~\cite{bosi2018computational, garcia2018sensitivity, masci2019impact, grigoriadis2020wall, feng2019analysis, garcia2021demonstration}, whereas other studies~\cite{ garcia2018sensitivity, masci2019impact, grigoriadis2020wall, feng2019analysis} investigated the influence of pulmonary vein configuration in LAA flow stasis.
The effect of atrial wall functionality has also been studied to investigate how it could contribute to thrombus formation~\cite{zhang2008characterizing, koizumi2015numerical, otani2016computational,bosi2018computational, masci2020proof}. For example, Koizumi et al.~\cite{koizumi2015numerical} and Masci et al.~\cite{masci2020proof} reported that decreased LAA contractility, often occurring in AF, can result in low velocity in the LAA and could affect LAA washout. 
Besides, some studies have also sought correlations between stroke and various in silico hemodynamic indices~\cite{ koizumi2015numerical, duenas2021comprehensive, paliwal2021presence, garcia2018sensitivity, aguado2019silico, gonzalo2022non}. Koizumi et al.~\cite{koizumi2015numerical} reported that increased RRT, as an indicator of thrombus formation, could contribute to flow stagnation in some parts of the LAA. Paliwal et al.~\cite{paliwal2021presence} reported  blood flow oscillations and elevated values of ECAP in fibrotic regions of the left atrial wall.

From the studies listed above, it is clear that CFD modelling of LA blood flows have reached an acceptable level of complexity by using dynamic wall motion, patient-specific boundary conditions, non-Newtonian rheology, and have been able to provide valuable insight, e.g., correlating plausible flows and stresses with disease states. 
However, as presented in Table~\ref{tab_summary_literature}, there is no consensus on how to model atrial flows, but there is also a dichotomy in methodological approaches, in terms of the specified spatial and temporal resolution, and solution accuracy. Most studies have used from 0.1 to 2k time-steps per cardiac cycle with 0.1 to 3M element meshes (cf. Table~\ref{tab_summary_literature}). On the other hand, there are only a few studies, in which 10 to 20k time-steps per cardiac cycle and meshes of 5 to 17M tetrahedral elements (or the equivalent) have been used.  
Still, the accuracy of the  solvers has rarely been reported, and no comprehensive assessment of the modelling choices and their impacts on the computed hemodynamic indices has been presented, to our knowledge. Therefore, the aim of the present study was to investigate the independent effects of spatial and temporal resolution, and combined effects of solution strategies, on the atrial flow patterns and their impacts on qualitative and quantitative hemodynamic indices. 

\section{MATERIALS AND METHODS}\label{sec_methods}

\subsection{Data acquisition}
Medical images were provided by Haut-L\'ev\^eque Hospital (Bordeaux, France), where cardiac computed tomography (CT) imaging was performed on a 64-slice dual source CT system (Siemens Definition, Siemens Medical Systems, Forchheim, Germany). The study was approved by the Institutional Ethics Committee, and all patients provided informed consent. Left atrial geometries were extracted as detailed in Mill et al.~\cite{mill2021silico} We focus here on a subset of 12 cases to minimise work, while simultaneously cover a representative breadth of morphologies found in a population. The variability of morphological phenotype was quantified based on LA and LAA volumes, number of pulmonary veins and different types of LAA morphology, subjectively classified by following the procedure in Di Biase et al.~\cite{di2012does}. The morphological characteristics of the cohort are summarized in Table~\ref{tab_cohort}. The segmented surfaces were minimally smoothed (cf. Figure~\ref{fig_generic_flowrate} (A)) by using \textit{MeshMixer}~\cite{web_meshmixer}. The pulmonary veins (PVs) and the mitral valve (MV) were extended with the length equivalent of their diameter, as shown in Figure~\ref{fig_generic_flowrate} (B), which also shows the distinction of different regions, where the LA and the LAA have been separated for post-processing purposes using objective and automated algorithms in \textit{morphMan}~\cite{kjeldsberg2019morphman}.

 \input{Table_1}

\subsection{Computational fluid dynamics}\label{sec_cfd_intro}

CFD simulations were performed using \textit{Oasis}~\cite{mortensen2015oasis}, an open-source library~\cite{web_oasis} solving the Navier-Stokes equations using the Finite Element method, based on the FEniCS~\cite{web_fenics} computing platform, which has been rigorously and successfully verified and validated~\cite{khan2019direct, bergersen2019fda,haley2021delayed}, \textit{Oasis} is a high-performance computing implementation of a segregated, space/time centered, incremental pressure correction scheme. The convection term is discretized using an Adams–Bashforth-projected convecting velocity and Crank–Nicolson is used for descretization of the convected velocity as well as diffusive term. These specific choices ensure overall second-order accuracy and a solution that preserves kinetic energy by minimizing numerical dispersion and diffusion~\cite{karniadakis2005spectral}, More information regarding the solver and numerical implementation can be found in the original manuscript~\cite{mortensen2015oasis}. We used Lagrange finite elements of order 1 ($\mathbb{P}_1$) for both the velocity and pressure, both of which are second-order accurate (L2 norm).

\subsection{Boundary conditions and modelling assumptions}

\begin{figure}[!htbp] 
\centering
\includegraphics[width=0.98\textwidth]{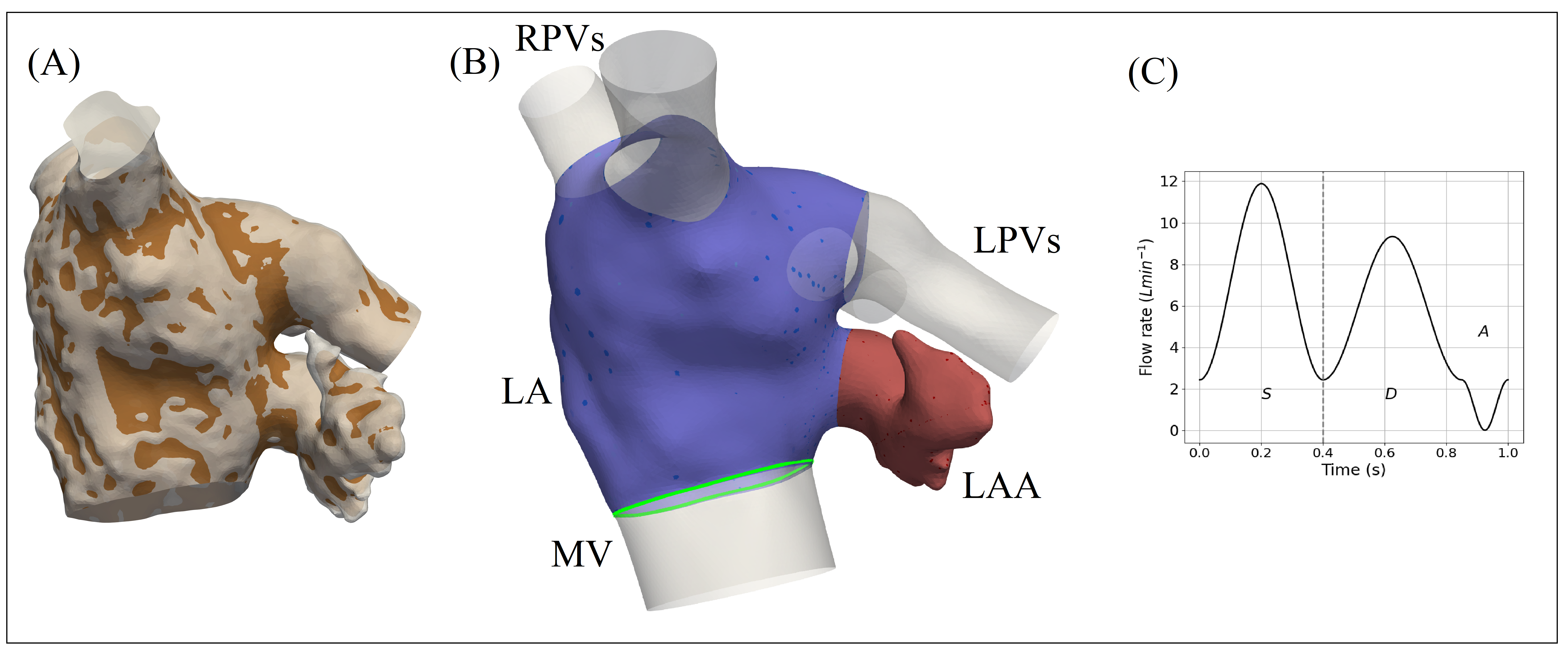}
\caption{(A) Original (gray) and smoothed (brown) left atrial geometry of case 167, (B) Left atrium (LA, blue), left atrial appendage (LAA, red), mitral valve (MV, green) and the right and left pulmonary veins (RPVs and LPVs, respectively). The applied flow extensions at PVs and MV are also visible. (C) Generic flow rate over one cardiac cycle. S, D and A stand for systole, diastole and atrial kick, respectively.}
\label{fig_generic_flowrate}
\end{figure}

We applied a generic waveform~\cite{smiseth1999pulmonary, fernandez2012analysis} (cf. Figure~\ref{fig_generic_flowrate} (C)) at the PVs, where the flow rate was scaled with respect to cross-sectional area and prescribed as parabolic velocity profile. We applied a normal cardiac output of 5.5 {Lmin}$^{-1}$~\cite{kumada1967cardiac, smiseth1999pulmonary}. We simulated two cardiac cycles, using the first cycle to minimise the effects of artificial initial conditions, and the second cycle for evaluating results. We assumed rigid walls and an open mitral valve, where the pressure was set to zero. At the outlet boundary, we also used a back-flow stabilization for the velocity~\cite{esmaily2011comparison}. Blood was modelled as an incompressible and Newtonian fluid with constant density of $\rho = 1060$ kg/m$^3$ and dynamic viscosity of $\mu=$0.0035 Pa*s.

\subsection{Mesh convergence study}
\begin{figure}[!htbp] 
\centering
\includegraphics[width=0.99\textwidth]{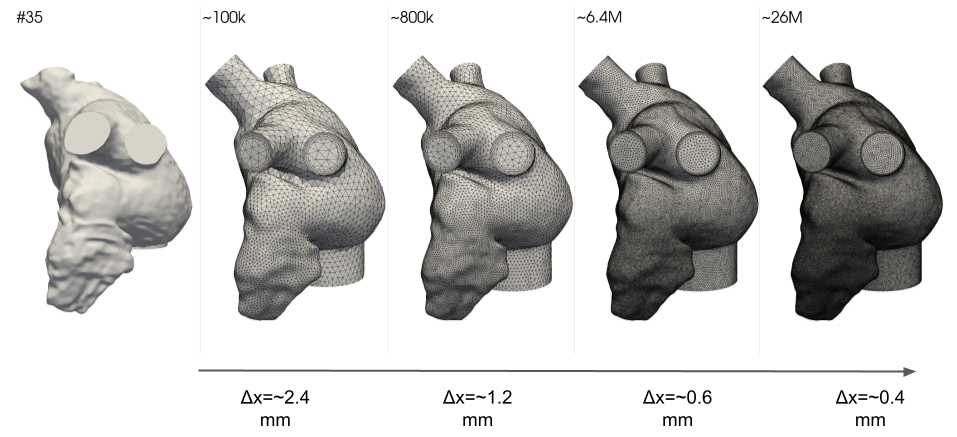}
\caption{The leftmost shows the initial model (case 35) before smoothing and mesh generation. On the right, it shows four different meshes; 100k, 800k, 6.4M, and 26M elements. Meshes with 400k and 3.2M elements are not presented for the sake of simplicity.}
\label{fig_4meshes_C35}
\end{figure}

We investigated the effect of spatial resolution on the 12 cases of the cohort by using six different meshes varying from 100k to 26M tetrahedral elements to cover the span of used mesh resolutions in the literature (cf. Table~\ref{tab_summary_literature}). We used an objective, automated, and robust open-source \textit{VMTK}-based~\cite{web_vmtk} meshing algorithm implemented in \textit{VaMPy}~\cite{web_vampy}. We specified a uniform characteristic edge length, which is automatically adjusted to capture high surface curvatures, especially in the LAA, as shown in Figure~\ref{fig_4meshes_C35}, using four boundary layers. Our strategy was to start with the coarsest possible mesh 100k elements (i.e., $\Delta x=\sim$2.4 mm) and dividing the element edge length by 2 successively, which in 3D increased the mesh sizes by a factor $2^{3}$. This resulted in 800k and 6.4M element meshes, but we  also considered  intermediate mesh sizes consisting of 400k and 3.2M elements. To establish a point of reference, we used 26M element mesh (i.e., $\Delta x=\sim$ 0.4 mm). The mesh characteristics are summarized in Table~\ref{tab_mesh_sizes}. The mesh convergence study was performed using 10,000 time-steps per cardiac cycle. 

 \input{Table_2}

\subsection{Time convergence study }
Following the results of the mesh resolution study, we investigated the sensitivity to temporal resolution on three selected cases that showed high, medium, and low sensitivity to mesh resolution (case number 4, 26, and 192). We used 6.4M element mesh and varied the time-steps per cardiac cycle from 1250, 2500, 5000, 10,000 to 20,000, which corresponds to $\Delta$t = 0.8 ms, 0.4 ms, 0.2 ms, 0.1 ms and 0.05 ms. The corresponding calculated range of Courant–Friedrichs–Lewy (CFL) numbers in the domain was CFL$\approx$ 0.4--0.025.

\subsection{Combined solution strategies}  
Following the mesh and time convergence studies of individual cases, we investigated the combined effects of spatial/temporal resolution, and solver accuracy on the entire cohort. We therefore compared the commonly used normal-resolution/accuracy (NR) simulations (median values as presented in Table~\ref{tab_summary_literature}) against the high-resolution (HR) ones to assess whether HR simulations would provide any added value statistically. 
To ensure the numerical experiment is as controlled as possible, a dedicated NR solver~\cite{web_nr_solver} was developed by using FEniCS~\cite{web_fenics} for NR simulations. Navier--Stokes equations were discretized in time and space by using implicit backward Euler, with streamline upwind Petrov-Galerkin (SUPG) stabilization scheme. 1000 time-steps per cardiac cycle and 800k elements (i.e., as presented in Table~\ref{tab_summary_literature}) were used for NR simulations (i.e., corresponding to CFL$\approx$ 0.3). \textit{Oasis} as introduced in Section~\ref{sec_cfd_intro} was used for HR simulations with 10,000 time-steps per cardiac cycle, and 26M element mesh (i.e., corresponding to CFL$\approx$ 0.07). 
The HR and NR solvers' order of accuracy were verified in space and time, as presented in Appendix~\ref{app_order}, Figure~\ref{fig_order_of_accuracy} (a) and (b), respectively.

 \input{Table_3}

\subsection{Post-processing} 
We computed the four most commonly computed hemodynamic indices; WSS, OSI, RRT, and ECAP. The mathematical formulations are presented in Table~\ref{tab_hi} of Appendix~\ref{HIs_definition}.

\section{RESULTS}\label{sec_results}

\subsection{Mesh convergence study}\label{subsec_mesh_refinement}
\begin{figure}[!htbp] 
    \centering
\includegraphics[angle=0, width=1.0\textwidth]{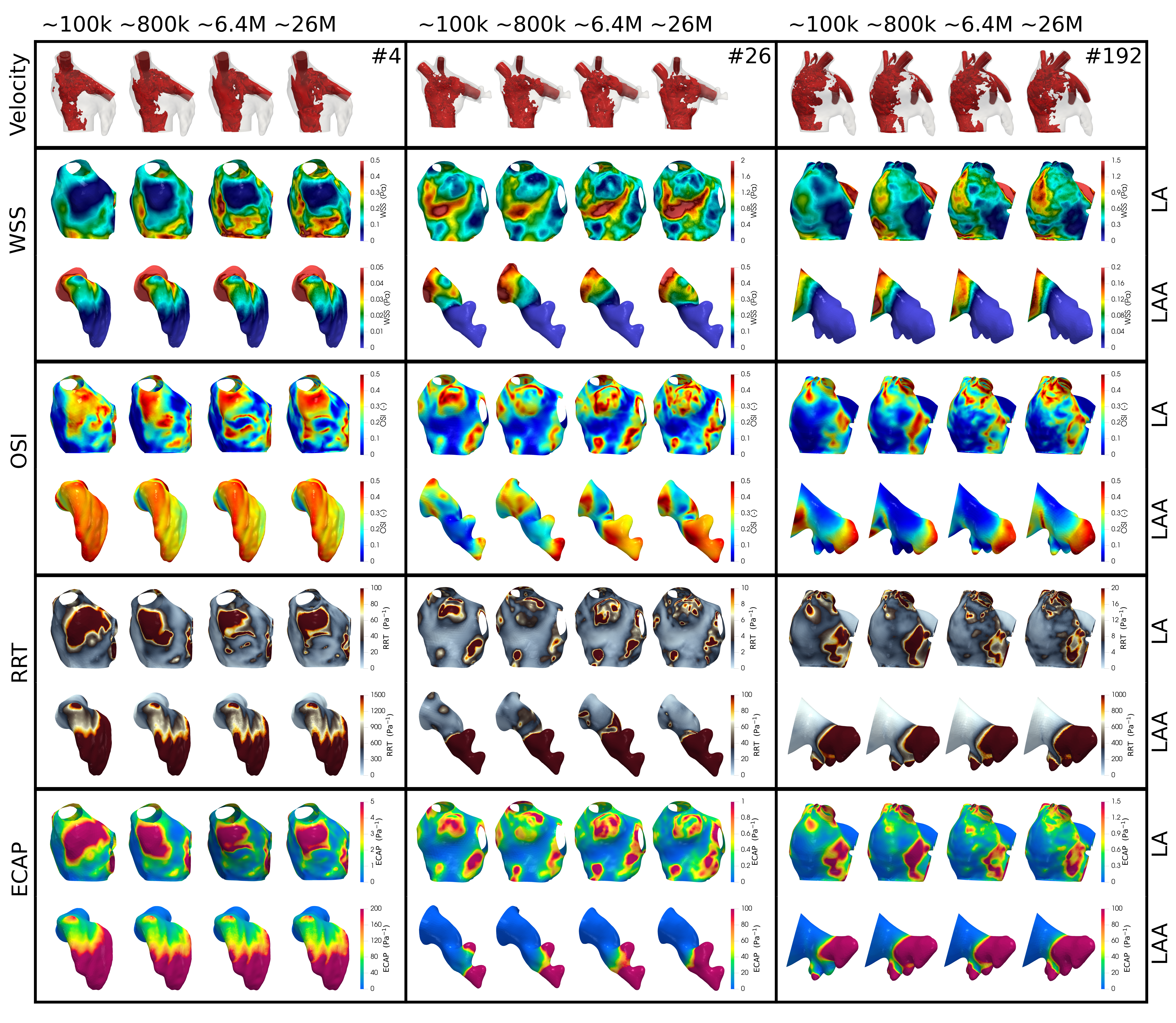}
\caption{Qualitative impact of mesh resolution (100k, 800k, 6.4M, and 26M elements) on cases 4, 26 and 192. Results of 400k and 3.2M element mesh are not presented for the sake of simplicity. For each case, the first row compares isovelocity surfaces (in the range of [0.18-0.22] m/s), the second and third rows compare time averaged wall shear stress (WSS) in the left atrium (LA) and left atrial appendage (LAA), respectively, the fourth and fifth rows show oscillatory shear index (OSI) in LA and LAA, respectively, sixth and seventh rows present relative residence time (RRT) and finally eighth and ninth rows present endothelial cell activation potential (ECAP) results.}
\label{fig_qualRes_3cases}
\end{figure}

\begin{figure}[!htbp] 
\centering
\includegraphics[width=0.68\textwidth]{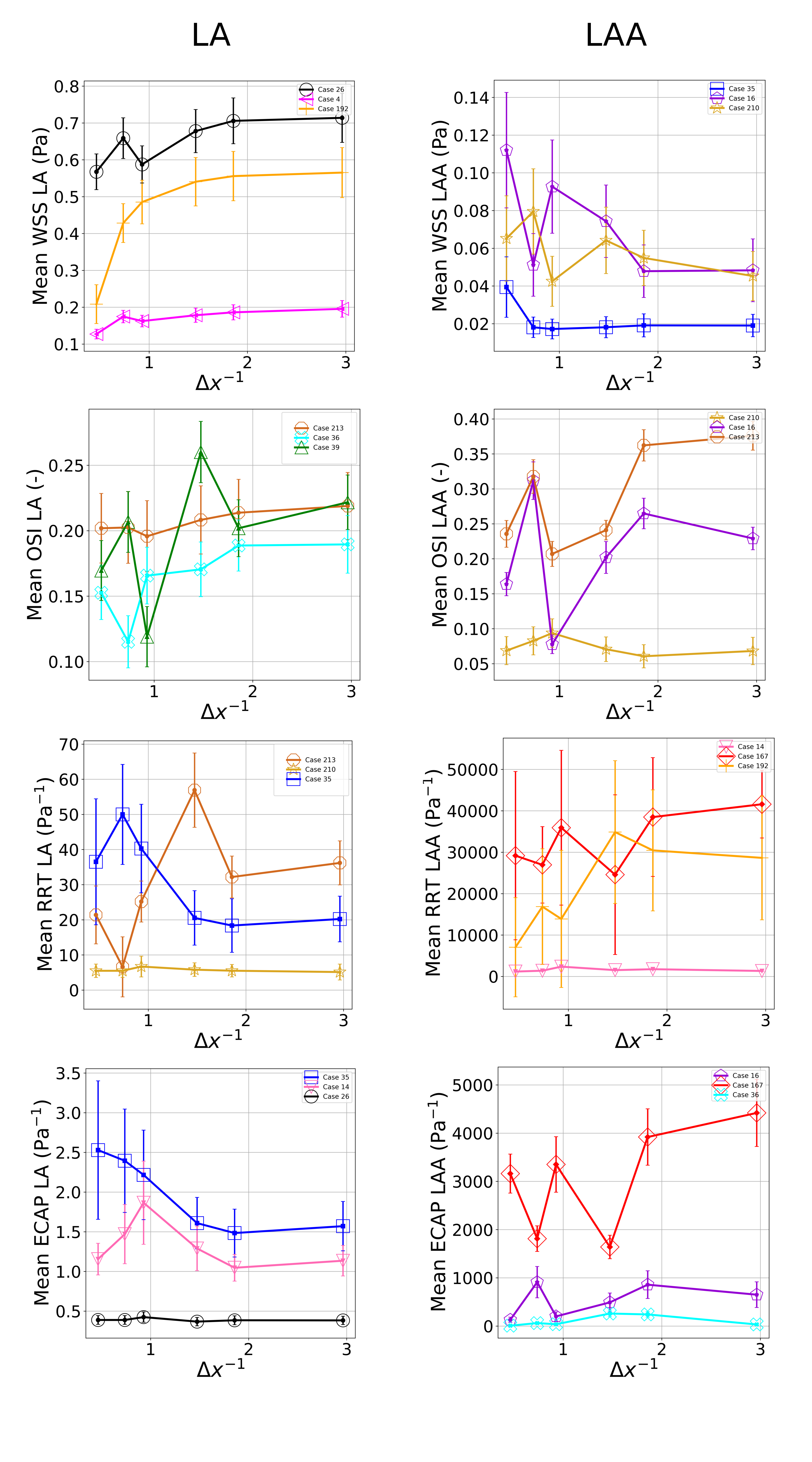}
\caption{Quantitative results of mesh convergence study of time averaged wall shear stress (WSS), oscillatory shear index (OSI), relative residence time (RRT), and endothelial cell activation potential (ECAP) in left atrium (LA) and left atrial appendage (LAA) separately. Representative cases of high, medium, and low sensitivity of each hemodynamic indices are separately presented for the LA and the LAA. $\Delta x=\sim2.4, \sim1.2, \sim0.6$, and $\sim0.4$ mm corresponds to meshes of 100k, 3.2M, 6.4, and 26M elements, respectively (cf. Table~\ref{tab_mesh_sizes}). Each color presents one case as shown in Figure~\ref{fig_all_results} Appendix~\ref{app_grid}.}
\label{fig_selected_mesh_refinement}
\end{figure}

First, we focus on qualitative results for three representative cases (i.e., number 4, 26, and 192), in which the WSS quantitatively showed low, medium, and high sensitivity to mesh resolution (cf. Figure~\ref{fig_selected_mesh_refinement}). These cases are also qualitatively presented for time convergence (Section~\ref{subsec_time_refine}) and combined effects (Section~\ref{subsec_hr_vs_nr}) to enable a thorough visual comparison among the results.
Figure~\ref{fig_qualRes_3cases} shows isovelocity surfaces (in the range of [0.18-0.22] m/s), WSS, OSI, RRT, and ECAP for 100k, 800k, 6.4M, and 26M element meshes, respectively. The hemodynamic indices are presented separately for LA and LAA. We can observe that there are clear discernible differences in isovelocity patterns on different mesh resolutions, as also presented in a zoomed-in version in Figure~\ref{fig_3_isocel_mesh} of Appendix~\ref{app_grid}. For the coarse meshes with 100k elements, the flow patterns are difficult to interpret and the flows are not smooth or can be reflective of numerical artifacts. In contrast, we can observe that the higher resolution meshes captures more detailed and complex flow structure.
Refining the mesh leads to phenotypically different WSS patterns with different regions with the highest values, especially for the meshes larger than 6.4M. These effects are relatively pronounced in the LAA for the regions near the ostium. By definition, OSI is arguably more sensitive to mesh resolution than WSS. By refining the mesh, different OSI patterns and high/low regions appear. Differences are less pronounced in the LAA, however, there are noticeable differences for case 26 for finer meshes compared to coarser ones. Similarly, RRT and ECAP show different patterns and high/low regions on the higher resolution meshes. Although it is difficult to visually observe the differences in the LAA compared to the LA, quantitative results (cf. Figure~\ref{fig_selected_mesh_refinement}) suggests that differences are still high in the LAA.

Quantitative results for all cases on all meshes are shown in Figure~\ref{fig_all_results} of Appendix~\ref{app_grid}. We identified three different phenotypical behaviours in the results, which we classified as: (1) High variability between mesh resolutions with a staggered pattern; (2) Lower variability but high errors between fine and coarse meshes; and (3) Inconsiderable/low variability with a smooth pattern. A subset of the representative quantitative results is shown in Figure~\ref{fig_selected_mesh_refinement}. The hemodynamic indices have been independently computed for the LA and the LAA. The average relative errors of all cases for WSS values between 800k element mesh (median value as shown in Table~\ref{tab_summary_literature}) and 26M element mesh (as a reference) are $16\pm34\%$ and $45\pm49\%$ in the LA and the LAA, respectively. However, we can see in the LA of case 192 (cf. top left in Figure.~\ref{fig_selected_mesh_refinement}), a WSS value of $0.21\pm0.02$ Pa on the 100k element mesh that increases to $0.57\pm0.05$ Pa on 26M element mesh, (i.e., 270\% increase, with a $63\%\pm22$ relative error), which classified as number (2). Moreover, there is a higher dependency of WSS to mesh resolutions in the LAA for cases 16 and 210. For case 16, WSS values show a staggered pattern as classification number (1), shifting from $0.05\pm0.01$ Pa to $0.09\pm0.02$ Pa and to a converged value of $0.04\pm0.02$ Pa, on the 400k, 800k, and 6.4M element mesh, respectively, which is 100\% variability. In contrast, case 26 shows classification number (3), where relative errors of WSS in the LAA stays below 5\% between meshes of 800k and 26M elements. 
The average relative errors of all cases for OSI are $12\pm0.1\%$ and $30\pm0.2\%$ in the LA and the LAA, respectively. However, we can see that OSI is more sensitive to mesh resolutions both in the LA and the LAA. Case 39 in the LA shows classification number (1), where the OSI value shifts from $0.12\pm0.018$ to $0.26\pm0.016$ and to $0.2\pm0.014$ on 800k, 3.2M and 6.4M element mesh, respectively. The OSI value in the LAA for case 213 also increases 170\% from $0.21\pm0.01$ to $0.37\pm0.01$ on the 800k and 26M element mesh, respectively.
RRT also shows high sensitivity to mesh resolution even though the average relative errors of all cases are $27\pm0.4\%$ and $74\pm0.6\%$ in the LA and LAA, respectively. For instance, it can be seen that for case 35 in the LA that classified as number (2), the RRT value $35.5\%\pm19$ Pa$^{-1}$ on the 100k element mesh, reduces to $20.22\pm8$ Pa$^{-1}$ on the 26M element mesh, (i.e., 180\% drop, with a $81\%\pm137$ relative error). For case 167, the RRT value in the LAA shows a staggered pattern with variability range of 10-80\% between the meshes of 100k to 26M elements.
Although ECAP shows relatively high sensitivity to mesh resolution particularly in the LAA, the average relative errors of all cases are $22\pm2.7\%$ and $66\pm0.5\%$ in the LA and the LAA, respectively. The ECAP value in the LAA for case 167 varies highly, shifting from $3327\pm440$ Pa$^{-1}$ to $1836\pm521$ Pa$^{-1}$, and $3904\pm0.02$ Pa$^{-1}$ on the 800k, 3.2M, and 6.4M element mesh, respectively. There is not a clear convergence of ECAP magnitudes for case 167 in the LAA, up to mesh 26M elements due to a high variance of differences $14\pm8\%-221\pm43\%$. 

To compare the results with some points of reference as presented in Section~\ref{subsec_relevance}, the averaged WSS value over all cases for 26M elements mesh is $0.44\pm0.12$ Pa and $0.036\pm0.06$ Pa in the LA and LAA, respectively. For OSI, the averaged value of all cases for 26M elements mesh is $0.21\pm0.11$ and $0.24\pm0.09$ in the LA and the LAA, respectively. And the averaged of normalized RRT and ECAP with respect to mean values for 26M elements are $0.93\pm0.21$ and $1.05\pm0.48$ in the LA, and $1.1\pm0.16$ and $0.92\pm0.52$ in the LAA.

\subsection{Time convergence study}\label{subsec_time_refine}

\begin{figure}[!htbp] 
\centering
\includegraphics[angle=0,width=1.0\textwidth]{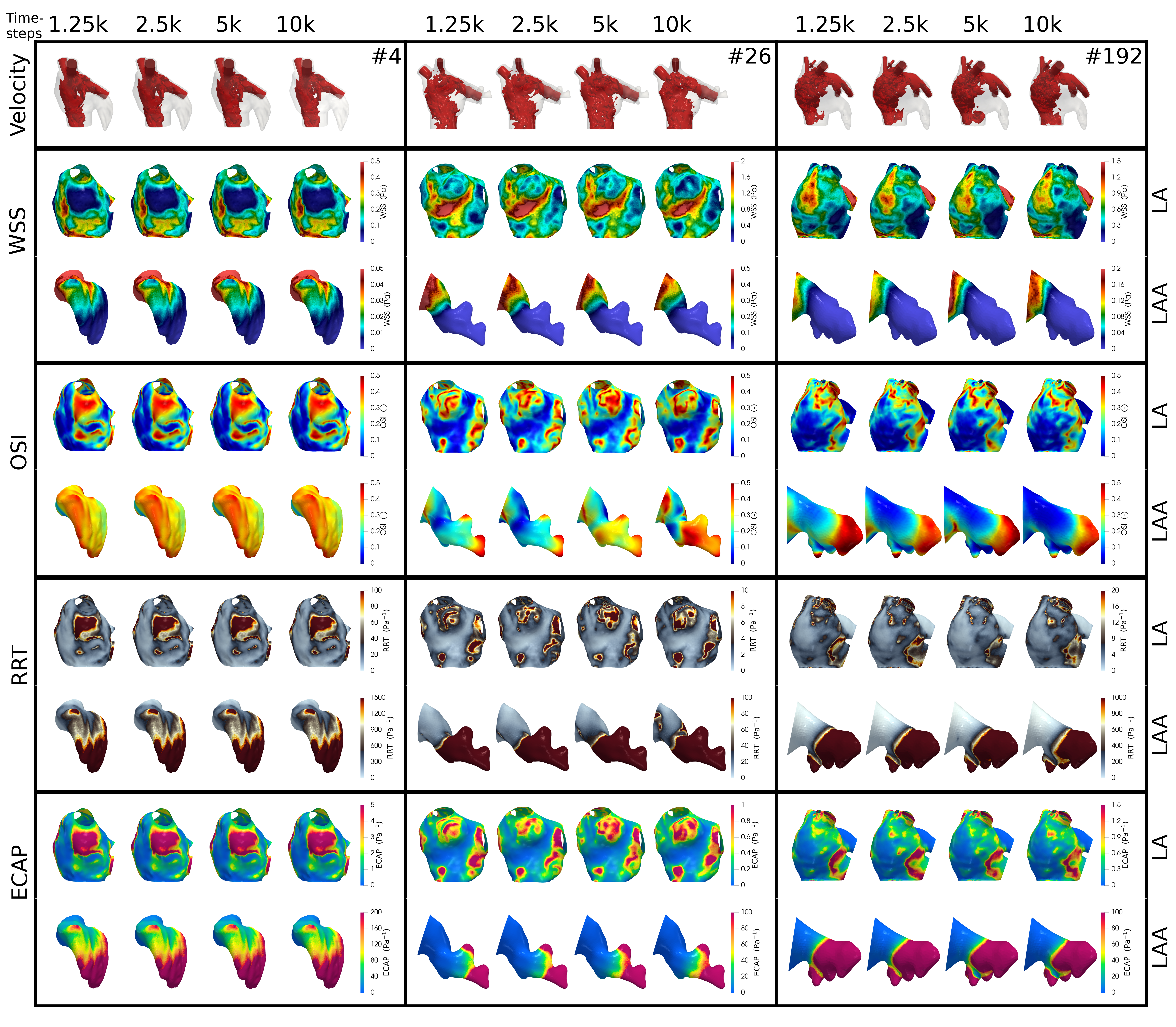}
\caption{Qualitative results at 1250, 2500, 5000 and 10,000 time-steps per cardiac cycle for cases 4, 26, and 192 in the LA and the LAA separately. For each case, the first row compares isovelocity surfaces (in the range of [0.18-0.22] m/s), the second and third rows compare time averaged wall shear stress (WSS), the fourth and fifth rows show oscillatory shear index (OSI), sixth and seventh rows present relative residence time (RRT) and finally eighth and ninth rows show endothelial cell activation potential (ECAP) in the left atrium (LA) and left atrial appendage (LAA) separately.}
\label{fig_qual_3cases_dts}
\end{figure}

\begin{figure}[!htbp] 
\centering
\includegraphics[width=0.68\textwidth]{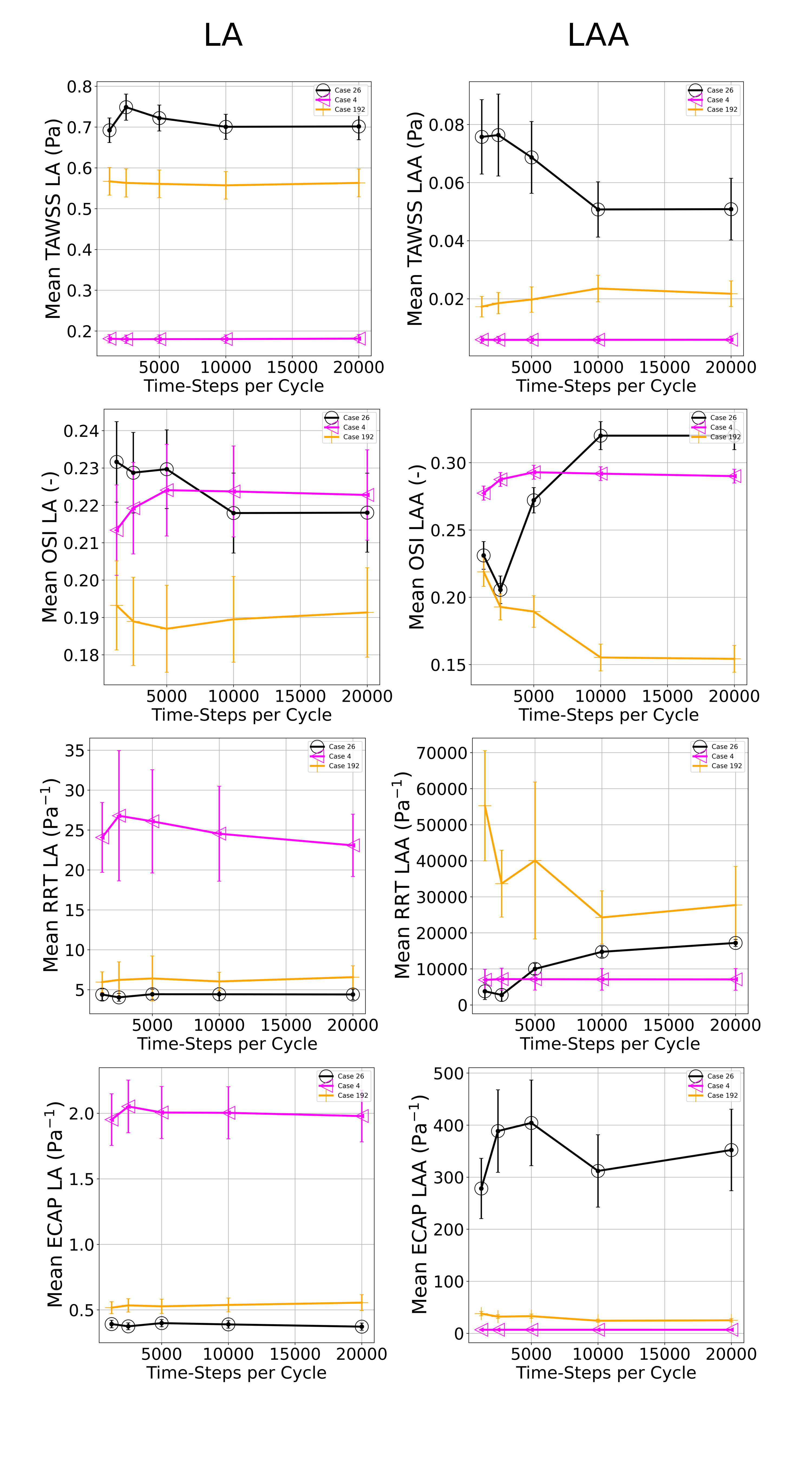}
\caption{Quantitative results of 
time averaged wall shear stress (WSS), oscillatory shear index (OSI), relative residence time (RRT), and endothelial cell activation potential (ECAP) in the left atrium (LA) and the left atrial appendage (LAA) for cases 4, 26 and 192 at 1250, 2500, 5000, 10,000 and 20,000 time-steps per cardiac cycle.}
\label{fig_quan_3cases_dts}
\end{figure}

Figure~\ref{fig_qual_3cases_dts} shows the qualitative results for isovelocity surfaces (in the range of [0.18-0.22] m/s), WSS, OSI, RRT, and ECAP in the LA and LAA separately. To concisely present the results, only simulations at 1250, 2500, 5000 and 10,000 time-steps per cardiac cycle are presented. Although there are small changes in flow patterns in cases 26 and 192 by refining temporal resolution, relatively indistinguishable flow changes can be seen for case 4. This is reflected on the WSS results as well, where case 26 and 192 show slightly different patterns in the LA, and amplified levels in the regions near ostium in the LAA. The latter suggests that fundamental flow changes occur in the LA, which affects the inflow to the LAA.
OSI also shows sensitivity in cases 26 and 192, where different patterns can be seen in the LA. Differences are noticeable in the LAA as well, especially for case 26.  
RRT and ECAP in the LA show different patterns in the region near the ostium where there is low WSS and high OSI. Although it is difficult to find the qualitative differences in the LAA, quantitative results (cf. Figure~\ref{fig_quan_3cases_dts}) suggest that there are high differences in RRT and ECAP in the LAA, especially for case 26.

The quantitative results of WSS, OSI, RRT, and ECAP in the LA and LAA for cases 4, 26, and 192 are separately presented in Figure~\ref{fig_quan_3cases_dts}. Although WSS values in the LA are robust to the temporal resolutions relatively for all three cases, WSS value in the LAA for case 26 decreases from $0.075\pm0.012$ Pa at 1250 time-steps/cycle to $0.05\pm0.1$ at 10,000 time-steps/cycle which is a 50\% decline, relative error of $50\pm0.2\%$. 
However, OSI shows more sensitivity particularly in the LAA, where for case 26, it shifts from $0.23\pm0.01$ to $0.2\pm0.01$, and to converged value of $0.32\pm0.01$, at 1250, 2500, and 20,000 time-steps/cycle, respectively, 30\% variability. However, overall variability below 5\% are generally obtained in the LA and LAA at time-steps larger than 5000 per cardiac cycle. Similar behavior can be seen for RRT and ECAP where variability exists noticeably in the LAA. For case 192, RRT values in the LAA varies roughly from $55269\pm15300$ at 5000 time-steps/cycle to $40078\pm21767$ at 10,000 time-steps/cycle, and to $27772\pm10171$ at 20,000 time-steps/cycle, indicating 200\% overall decrease. ECAP values in the LAA for case 26 show 30\% variability, shifting from $404\pm85$ at 5000 time-steps/cycle to $311\pm70$ at 10,000 time-steps/cycle to even $352\pm75$ at 20,000 time-steps/cycle, relative error of $12\pm0.1\%$.

\subsection{Combined solution strategies}\label{subsec_hr_vs_nr}

\begin{figure}[!htbp] 
\centering
\includegraphics[width=0.8\textwidth]{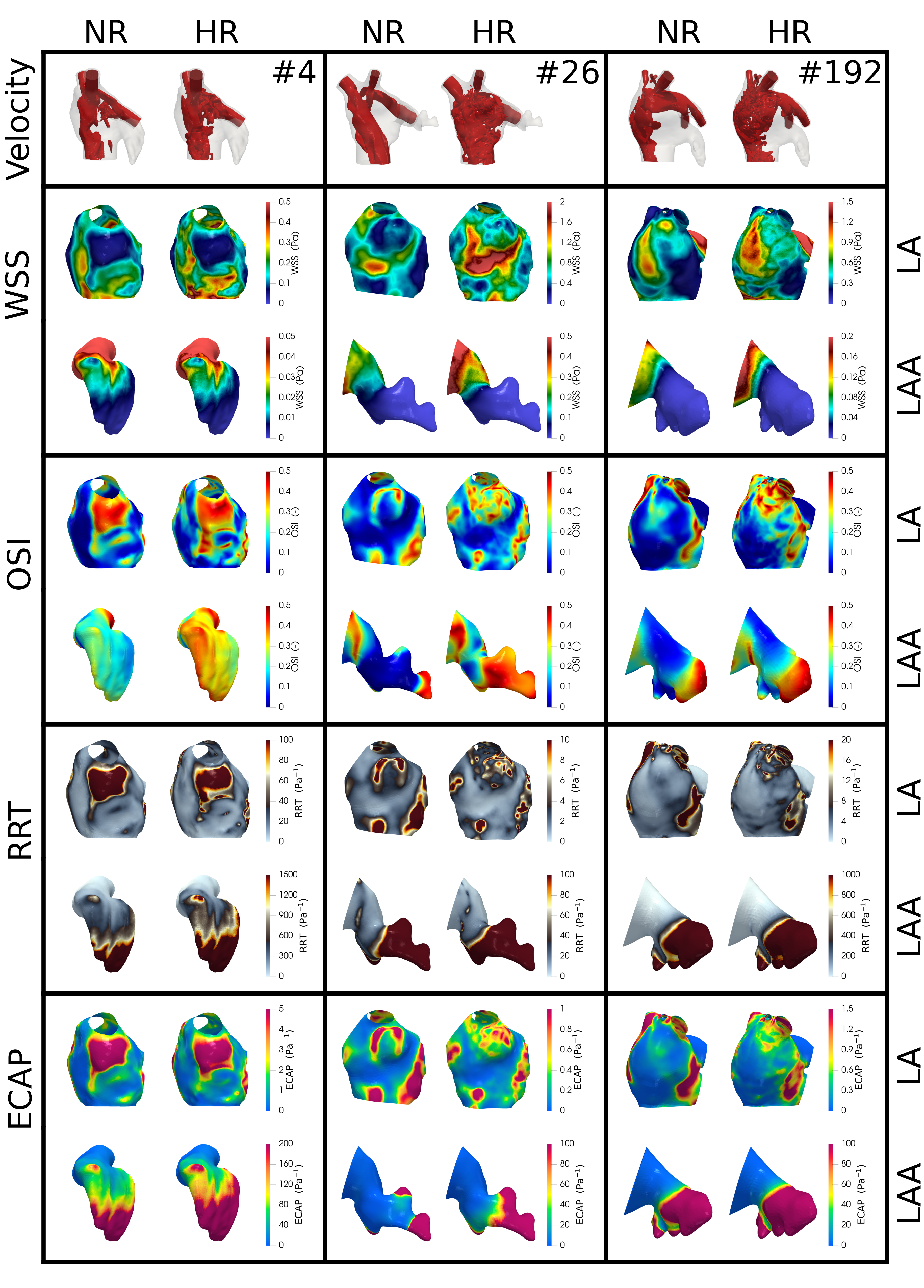}
\caption{Qualitative results of HR and NR simulations for cases 4, 26, and 192 in the left atrium (LA) and the left atrial appendage (LAA) separately. For each case, the first row compares isovelocity surfaces (in the range of [0.18-0.22] m/s), the second and third rows compare time averaged wall shear stress (WSS), the fourth and fifth rows show oscillatory shear index (OSI), sixth and seventh rows present relative residence time (RRT) and finally eighth and ninth rows present endothelial cell activation potential (ECAP).}
\label{fig_qual_hr_vs_nr}
\end{figure}

Figure~\ref{fig_qual_hr_vs_nr} shows isovelocity surfaces (in the range of [0.18-0.22] m/s) and maps of WSS, OSI, RRT, and ECAP in the LA and LAA separately, for HR and NR simulations. All NR simulations show smoother flows in contrast to the HR simulations, which show complex flows with fine structures. This can be seen clearly for cases 26 and 192, where flows have fundamentally different phenotypes. 
HR simulations show different WSS patterns and high/low regions compared to the NR ones in the LA, the differences in the LAA can mainly be observed in the regions near the ostium. HR simulations also predict different OSI patterns both in the LA and LAA compared to NR, which is in alignment with the observed different flow behaviours. 
The differences in RRT and ECAP are obvious in the LA for all three cases. However, it needs a closer look to observe the differences in the LAA, especially for RRT. 

\begin{figure}[!htbp] 
\centering
\includegraphics[width=0.61\textwidth]{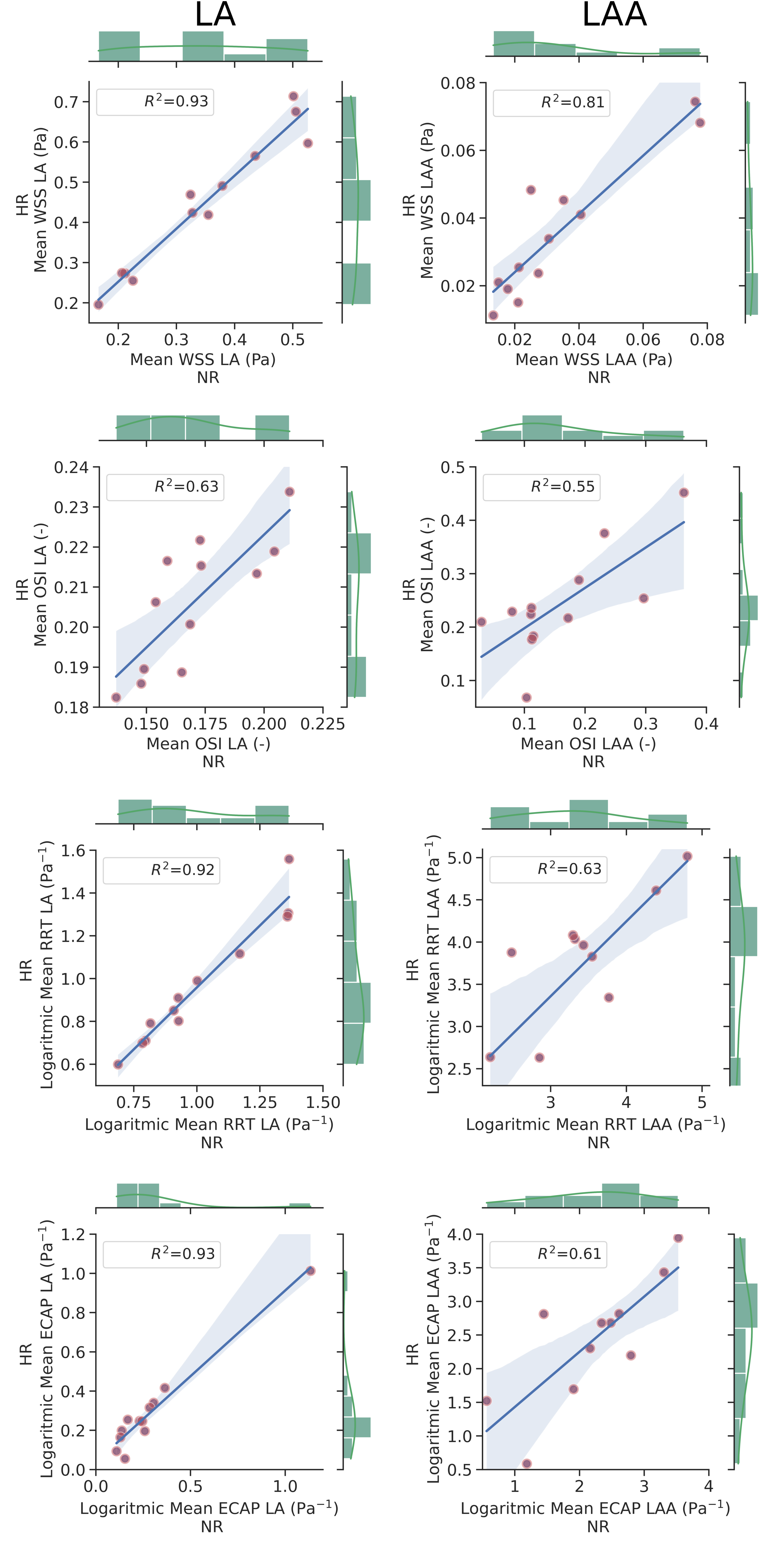}
\caption{Quantitative comparison of high- (HR) versus normal-resolution/accuracy (NR) simulations for all cases. From top to bottom: time averaged wall shear stress (WSS), oscillatory shear index (OSI), relative residence time (RRT), and endothelial cell activation potential (ECAP), for left atrium (LA) on the left side and left atrial appendage (LAA) on the right side separately. $R^2$ coefficient of determination as the measure of correlation is presented for each of them. The light blue shade indicates the 95\% confidence level. The distribution of NR and HR hemodynamic indices as well as their normal distribution are shown on the top side and right side of each plot, respectively. }
\label{fig_quan_hr_vs_nr}
\end{figure}

Figure~\ref{fig_quan_hr_vs_nr} highlights the quantitative results/statistics between NR and HR simulations for WSS, OSI, RRT, and ECAP in the LA and LAA separately. WSS values in the LA and LAA show a robust correlation (i.e., $R^{2}$=0.93 and 0.81, respectively), averaged relative errors of all cases are $21\%\pm 6\%$ and $14\%\pm 11\%$, respectively. Although WSS patterns are qualitatively different in Figure~\ref{fig_qual_hr_vs_nr}, the domain-averaged values show relatively small difference.
OSI in both LA and LAA was highly affected and highlights a poor correlation (i.e., $R^{2}$=0.63 and 0.55, respectively), particularly in the LAA where the averaged relative error of all cases is $44\%\pm21\%$. The RRT and ECAP indicate poor correlation in the LAA as well (i.e., $R^{2}$=0.63 and 0.61, respectively) in contrast to the LA. The averaged relative errors of all cases in the LAA are $64\%\pm18\%$ and $56\%\pm23\%$. Note that a logarithmic scale for RRT and ECAP is used in Figure~\ref{fig_quan_hr_vs_nr}.

\section{DISCUSSION}\label{sec_discussions} 
We systematically assessed the effect of spatial and temporal resolution on atrial flows, and demonstrated that spatial resolution has a profound impact on flow patterns and commonly computed hemodynamic indices, more so than temporal resolution when using a space/time-centered second-order accurate solver. We observed minor differences in most cases between the results on the 6.4M elements and those of 26M elements, and similarly minor differences between 10,000 and 20,000 time-steps per cardiac cycle. We subsequently estimate that simulations on meshes of $\sim$10M elements and 10,000 time-steps/cycle would provide a sufficiently robust numerical result. Secondly, the HR and NR solution strategies resulted in vastly different flow phenotype and consequently hemodynamics patterns, particularly in the LAA. Of note is that the independent impact of spatial/temporal resolution seem higher than combined effects, but this is because we used the median value of mesh and time-steps/cycle (cf. Table~\ref{tab_summary_literature}) for the NR simulations. More importantly, we found a great inconsistency and variability in convergence rates, which shows that assessment of a single case and a single metric might not reflect the true sensitivity to modelling choices. In the following, we will discuss the relevance, relation to others, and strengths and limitations of our results.

\subsection{Relevance}\label{subsec_relevance}
In the majority of CFD studies of atrial flows, the authors have sought correlations between hemodynamics and disease~\cite{koizumi2015numerical,paliwal2021presence,grigoriadis2020wall}. It is generally difficult to contextualize the results of our study, as much of the analyses have been qualitative, and there is generally a lack of quantitative data in the literature. However, if we set aside all modelling variabilities, there are some points of reference that we can compare our results against. More specifically, Koizumi et al.~\cite{koizumi2015numerical} investigated the lack of atrial kick on hemodynamic stresses in both healthy and diseased cases. They highlighted approximately 60\% increase of OSI in the LAA from 0.12 in the healthy to 0.2 in the AF case because of high-frequency fibrillation of the atrial wall. 
Paliwal et al.~\cite{paliwal2021presence} showed that the average values of ECAP and RRT over eight cases for fibrotic regions are higher compared to the non-fibrotic regions. Normalized RRT and ECAP with respect to mean values are $1.22\pm0.11$ and $1.21\pm0.10$ on fibrotic regions, compared to $0.49\pm0.16$ and $0.5\pm0.15$ on non-fibrotic regions representing 58\% and 59\% relative errors, respectively. 
Grigoriadis et al.~\cite{grigoriadis2020wall} reported that WSS in the LA decreased up to 70\% with the abnormal velocity profile opposed to normal one; for instance in one case, WSS dropped from 0.54 Pa to 0.31 Pa. 

If we compare averaged errors over all 12 cases in the LAA (i.e., 44\%, 64\%, and 56\% for OSI, RRT, and ECAP) as presented in details in Section~\ref{subsec_mesh_refinement} and Section~\ref{subsec_hr_vs_nr}, against the previously reported correlations between hemodynamics and disease, the effects of numerical methods seem similar.  
However, if we investigate the individual cases, the sensitivity to modelling choices (or errors) are appreciable. For instance, WSS values become approximately 200\% larger for intermediate to finer meshes in the LA for case 192, more examples are reported in Section~\ref{subsec_mesh_refinement} and Section~\ref{subsec_time_refine}. 
Therefore, rank ordering of hemodynamic indices on intermediate spatial and temporal resolutions could lead to different conclusions compared to that of converged ones. Secondly, our results reveal that spatial/temporal convergence depends on the cases (i.e., on their geometrical and flow pattern complexity) as shown in Figures~\ref{fig_selected_mesh_refinement} and~\ref{fig_quan_3cases_dts}; for instance, some cases show smooth convergence where 1M elements is sufficient, in contrast to other cases with huge variability up to 26M elements. Going forward, this seems to leave us with two options; either to perform convergence study for each specific case or run all models at sufficiently robust resolution with an HR solver, where the latter seems to involve the least amount of man-hours. That being said, computational time is certainly increased for higher resolution meshes, but most state-of-the-art CFD software have excellent weak scaling capabilities on high performance computing clusters, and one can obtain nearly constant wall-clock times for various mesh sizes with many CFD solvers. 

\subsection{Relation to others}
Needless to say, we are not the first ones to address mesh convergence although our results are very different from previous results. Mesh convergence studies have previously only reported differences below five percent, and in contrast to us, concluded that normal spatial resolution is sufficient~\cite{otani2016computational,dahl2012impact, grigoriadis2020wall, duenas2021boundary}. Although we cannot pinpoint the precise cause, there are distinct methodological approaches. The previous range of the mesh sizes seem narrow compared to ours, typically only spanning a few hundred thousand elements, and not orders of magnitude as we have done. Secondly, in contrast to our 12 cases, only few selected ones have been considered, which might have masked the variability of sensitivities observed across our cases. Thirdly, only few quantities of interest (i.e., velocity, pressure, or hemodynamic indices) have been investigated, which again might not have been the most sensitive ones. Also special care must be taken to minimise numerical diffusion and dispersion errors\cite{roache1986editorial,valen2014mind} such as we have done, but such information is rarely reported.
Again, we do not know precisely what is being used in the literature, but as shown in Figure~\ref{fig_order_of_accuracy} of Appendix~\ref{app_order}, the convergence rate of first-order accurate (i.e., NR) solver is comparatively leveled with small changes in variables, particularly in a narrow range. The second-order accurate solver exhibits steeper convergence rates where differences in, e.g., mesh resolution will be more apparent on the flow and stress metrics. 

\begin{figure}[!htbp] 
\centering
\includegraphics[width=0.95\textwidth]{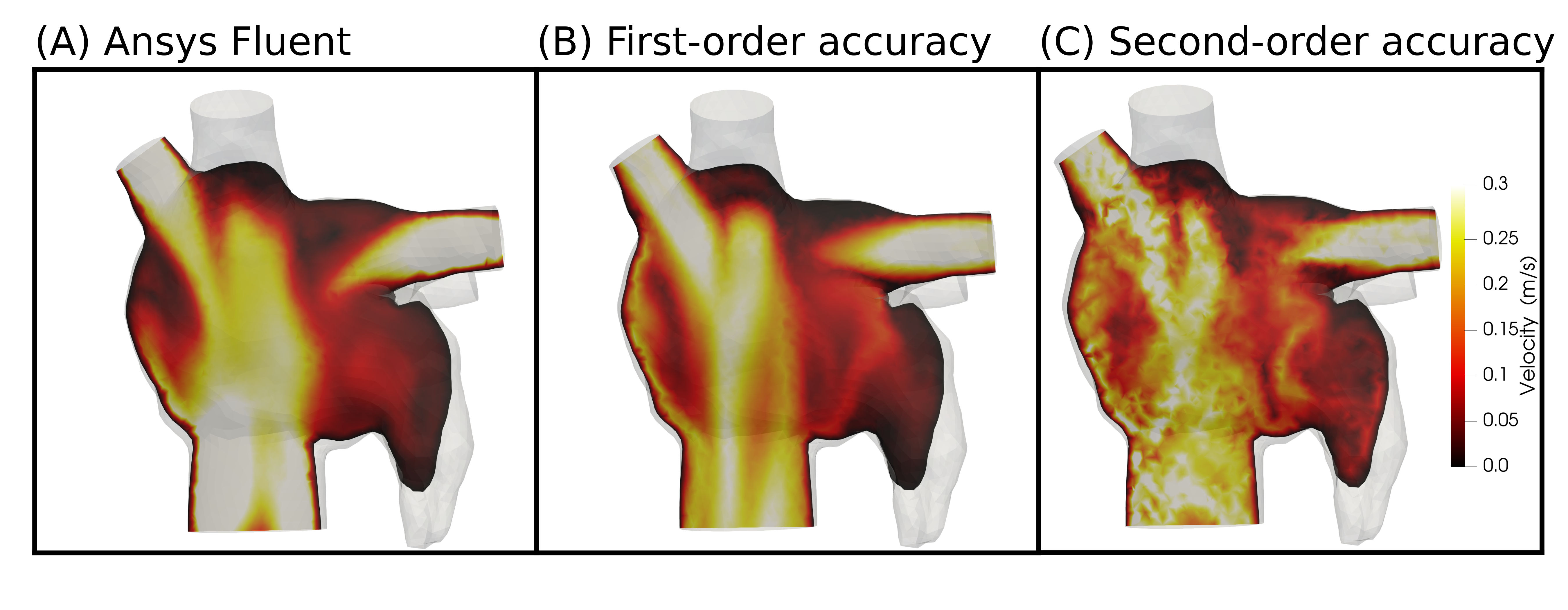}
\caption{Qualitative impact of the computational fluid dynamics solvers i.e., (A) Ansys Fluent, (B) First-order accuracy and (C) Second-order accuracy, on transient simulation of steady flow for case 35, with 150k element mesh and 100 time-steps/second. }
\label{fig_ansys_nr_hr}
\end{figure}

That being said, whether our NR solver is representative of the literature remains unknown. We therefore compared the solution of our NR solver against one of the most used commercial CFD solver, Ansys Fluent~\cite{web_ansys} on coarse spatial and temporal resolution, for the sake of simplicity. We chose case 35 and specified constant peak systolic velocity of 0.3 m/s (cf. Figure~\ref{fig_generic_flowrate}(C)). Simulations were performed for 2 seconds with 100 time-steps per second, and a 150k element mesh as commonly used in the literature (cf. Table~\ref{tab_summary_literature}). The Ansys Fluent~\cite{web_ansys} simulation was performed using the default settings of version 19.2. More specifically, the SIMPLE pressure-velocity coupling scheme, Green-Gauss scheme for gradients, and second-order upwind for pressure, momentum and transient formulation. The results, shown in Figure~\ref{fig_ansys_nr_hr}, indicate a strong sensitivity of flow dynamics to solver settings/accuracy, as expected. Ansys Fluent~\cite{web_ansys} performed somewhat similar to our NR solver, both indicating damping of flow instabilities predicted by HR solver. The flow phenotype was predicted to be smooth and laminar by Ansys Fluent~\cite{web_ansys} and NR solver in contrast to HR solver, where the flow was predicted to be highly unstable and transitional. 

\begin{figure}[!htbp] 
\centering
\includegraphics[width=0.8\textwidth]{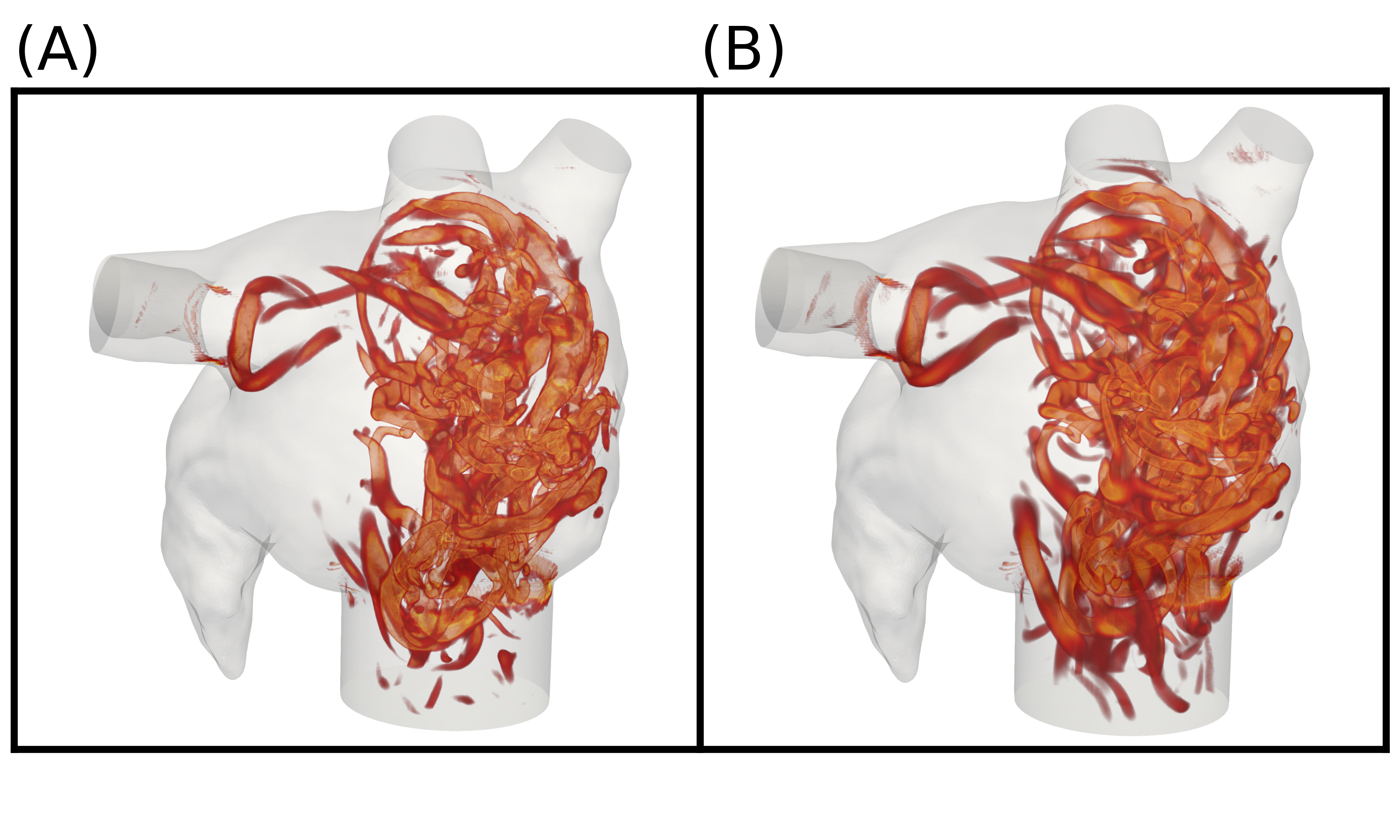}
\caption{The vortex patterns inside the left atrium are presented using Q-criterion (Q=0.002 s$^{-1}$), a posterior view at the end of systole (A) 26M elements, (B) 52M element mesh using $Oasis$.}
\label{fig_52M_c_35}
\end{figure}

In the current study we have taken a brute-force approach making no assumptions regarding the flow phenotype, in contrast to the vast majority of studies where the assumption of laminar flow has been explicitly stated. If we exclude Masci et al.\cite{masci2019impact, masci2020proof} who performed large eddy simulations, the most cited studies to date make this assumption explicitly. For instance, Otani et al.~\cite{otani2016computational} justified the assumption of modelling the flow as laminar because Reynolds number was below 2300, whereas Bosi et al.~\cite{bosi2018computational} reported it as a "substantial computational simplification". If we instead visualise our results, the Q-criterion of HR simulation for case 35 illustrated in Figure~\ref{fig_52M_c_35}(A) shows the presence of fine scale vortices. Qualitatively, this figure suggests that the flow is not strictly laminar. Secondly, our strategy for using mesh size of 26M elements was to go slightly beyond what has been reported in the literature to establish a point of reference, where numerical errors associated with spatial resolution is presumably zero. However, we have no guarantee that it is actually the case. We therefore doubled the mesh size for case 35, to a 52M elements as shown in Figure~\ref{fig_52M_c_35}(B). The differences in the scales of vortical structures between the 26M and 52M element mesh can be clearly seen, but there are negligible difference in e.g, WSS, where mean values are $0.149\pm0.095$ and $0.151\pm0.097$, respectively. The added value of a simulation on 52M element mesh remains an open question, more specifically, it can be questioned whether we now address accuracy or precision, given all the modelling assumptions associated with medical image-based CFD~\cite{steinman2019patient}.

\begin{figure}[!htbp] 
\centering
\includegraphics[width=0.51\textwidth]{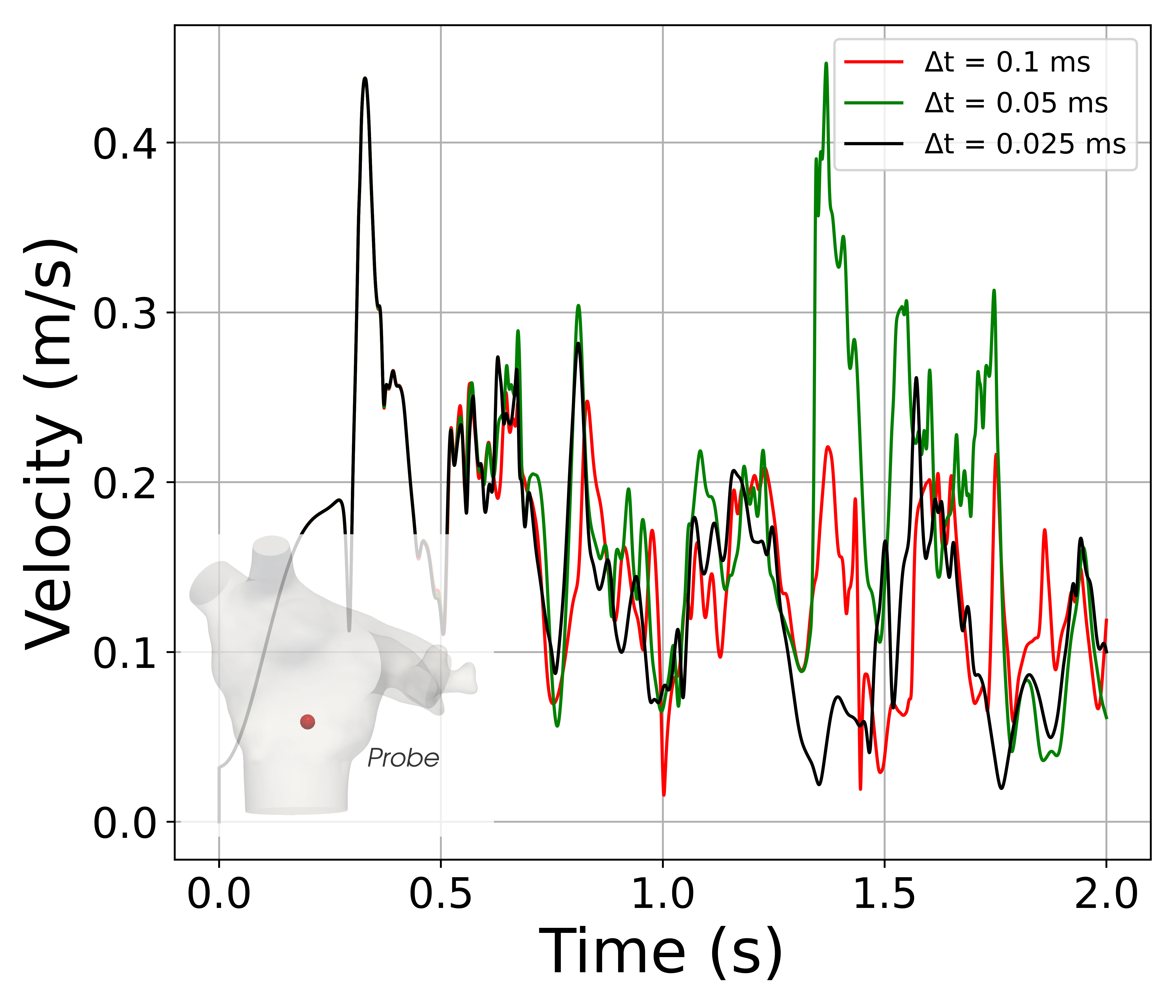}
\caption{Velocity profile at the probe point for three temporal resolutions 10,000, 20,000, and 40,000 time-steps/cycle. The location of the probe point is shown in the lower left corner.}
\label{fig_probe_5_c_26}
\end{figure}

Similarly, our conclusion was that 10,000 time-steps is sufficient to obtain converged hemodynamic indices in time. However, this is not necessarily true for the flow itself. Figure~\ref{fig_probe_5_c_26} displays the instantaneous velocity magnitude from a probe point for 10,000, 20,000, and 40,0000 time-steps/cycle simulations. Higher instantaneous fluctuations and velocity differences can be observed for 20k, and 40k time-steps/cycle compared to 10k time-steps/cycle. Therefore, the flow complexity of Q-criterion for 52M element mesh in Figure~\ref{fig_52M_c_35} and velocity fluctuations at 20k and 40k temporal resolution in Figure~\ref{fig_probe_5_c_26} indicate again that the assumption of laminar flow might need to be reconsidered in future studies.

\subsection{Strengths and Limitations}

Relative to the other studies ours is focusing on numerical and fluid mechanical aspects, rather than physiological ones; whether the observed flows are physical versus physiological can be debated. Whether these flows occur in the body remains an open question. What we can authoritatively say is that high-resolution/fidelity numerics are more accurate, compared to the lower resolutions and NR simulations. Moreover, our NR simulations do not reflect the range of solution strategies used in the literature (cf. Table~\ref{tab_summary_literature}), many studies have reported using 10 times coarser spatial and temporal resolution, which is 100-fold coarser combined. Hence, the comparison between HR and NR simulations is arguably conservative. 
There are also similarities between this study and previous works~\cite{valen2014mind,khan2015narrowing}. However, verification and validation is a moving target and it has been argued that addressing mesh convergence is required for every simulation~\cite{roache1986editorial}. Secondly, the conclusions here are somewhat different. For atrial flows, it seems that both the volume and Reynolds number are larger compared to aneurysms, which demands higher spatial resolution. 

We have not addressed the entire CFD modelling pipeline, which has shown to be highly sensitive to factors such as segmentation differences, as demonstrated for other cardiovascular regions~\cite{valen2018real}, and it is clear that operator dependent tasks such as semi-automatic LA geometry extraction from images can have profound impact on simulation results~\cite{berg2018multiple}. However, assessing such errors is beyond the scope of the current study.
Additionally, we have performed simulations for two cardiac cycles. However, NR solution strategies normally exhibit convergence within the first few cycles~\cite{otani2016computational,bosi2018computational,feng2019analysis},
whereas our HR solution strategy predicted transitional flow with random fluctuations in time and space. Such transitional flows are presumably equally different on every single cycle~\cite{garcia2021demonstration}, so the second cycle that we have used is only a representative random one. However, we admit that the number of beats required to converge the various hemodynamic indices is an open question. 
We have also assumed Newtonian rheology for blood, which is presumably valid in the LA where the high shear rates justifies this assumption. This is a common assumption in the literature and only two studies~\cite{wang2020numerical, gonzalo2022non} have considered non-Newtonian rheology for atrial blood flow. That being said, non-Newtonian effects could be more pronounced in the LAA where shear rates are expected to be an order of magnitude lower. 
We also assumed rigid walls, which is clearly unphysiological and an obvious limitation~\cite{mill2022computational}; nonetheless, this assumption is consistent with approximately half of the literature. Moreover, the left atrial cyclic volume change in AF patients is only 10-15\%~\cite{wozakowska2005changes, masci2020proof}, which represents a relatively minor movement of LA wall. Therefore, this limitation is not expected to effect the conclusion of our study. %
Since patient-specific flow rates are rarely accessible, we have used a generic waveform flow. Furthermore, we assumed that the flow rate is scaled with the cross-sectional area of PVs, implying the same inlet velocity i.e., $Q \sim D^{2}$. However, flow rate is typically assumed in the literature to split equally~\cite{lantz2019impact, garcia2018sensitivity, gonzalo2022non}, which specifies the same amount of flow (i.e., $Q \sim D^0$), and consequently very high velocity for the LA with many and small veins. 
We also assumed an open mitral valve, which is presumably valid in a convection dominated flow. 
Another limitation of our work is that we have focused on hemodynamic indices defined at the wall. This particular choice was made since they are well-defined and commonly computed in the literature. Nevertheless, caution must be taken as they are ultimately only surrogates of the flow. On the other hand, time-dependent volumetric indices such as Lagrangian particle tracking, and blood residence time, have more variable definitions~\cite{gonzalo2022non, d2020simulation} and are less frequently computed. It is possible that they could potentially be more sensitive to modelling strategies, and also better predictors of thrombus formation, but the latter remains to be demonstrated. In addition, the choice of metrics to compare simulation results is not obvious, and should be related to the question of interest and context of use of the computational model, following the terminology of the V\&V40 guidelines~\cite{v2018assessing}.

\section{CONCLUSION}\label{sec_conclusions}
In this study, we have presented a sensitivity analysis of spatial resolution, temporal resolution, and solver accuracy to assess the importance of modelling choices for predicting flows and hemodynamic indices in the left atria. It seems that there is indeed more than meets the eye. More specifically, there seems to be a profound sensitivity of modelling choices on predicting atrial flows, but also rank ordering of metrics, even on intermediate spatial/temporal resolutions, which could impact conclusions, depending on the question of interest and the context of use. That being said, we fully acknowledge that "all models are wrong"~\cite{box1979all}. and ours is no exception. Still, we believe that attention to fundamental aspects of CFD might be beneficial towards establishing a plausible model for atrial flows.

\section{ACKNOWLEDGEMENT}
This work was supported by the SimCardioTest project (Digital transformation in Health and Care SC1-DTH-06-2020) under grant agreement No. 101016496 and ERACoSysMed PARIS project under grant agreements No. 643271. 
The simulations were performed on the Saga cluster, with resources provided by UNINETT Sigma2 – the National Infrastructure for High Performance Computing and Data Storage in Norway, grant number nn9249k.


\bibliographystyle{unsrt}
\bibliography{scibib}

\clearpage

\appendix

\section{Accuracy of solvers}\label{app_order}

\begin{figure}[!htbp]
     \begin{center}
        \subfigure[]{%
            \label{fig:testcase1}
            \includegraphics[width=0.48\textwidth]{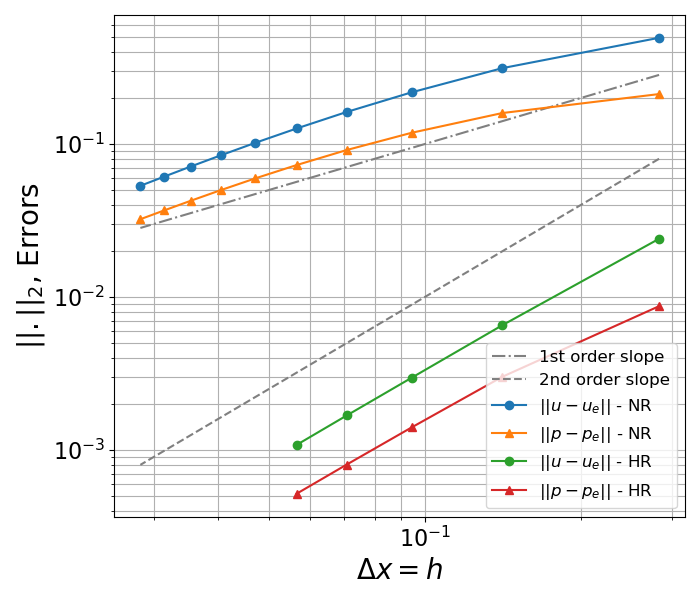} 
         } 
        \subfigure[]{%
            \label{fig:testcase2}
            \includegraphics[width=0.48\textwidth]{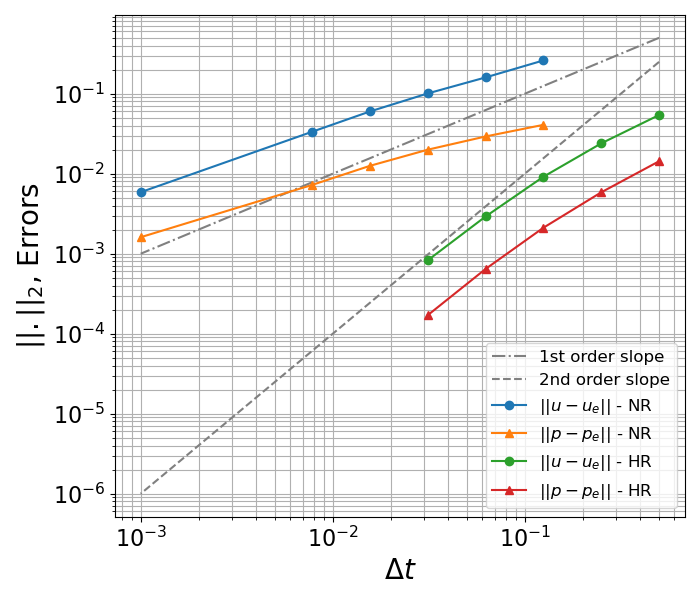}
 
        } 
    \end{center}
    \caption{%
        (a) $\text{L2}$ norm of errors of the velocity and pressure computed at various mesh levels for Taylor-Green problem. (b) $\text{L2}$ norm of errors of the velocity and pressure computed at various time-steps for Taylor-Green problem.
     }%
   \label{fig_order_of_accuracy}
\end{figure}

Since two dimensional Taylor–Green flow is one of the analytical and transient solutions to the Navier–Stokes
equations~\cite{mortensen2015oasis}, it was used for verification of the solvers i.e., $Oasis$ as high resolution (HR) and normal resolution (NR) solvers. The norms of the relative errors of the variables are indicators of the accuracy of the fluid solver when the mesh size and time-step change. The $\text{L2}$ norm of errors of velocity $\textbf{u}$ and pressure $p$ against analytical solutions were computed i.e., $ \lVert \textbf{u} - \textbf{u}_e \rVert_{2} $ and $ \lVert p - {p_e} \rVert_{2} $. The $\text{L2}$ norm of velocity and pressure are plotted in Figure~\ref{fig_order_of_accuracy}. 
To study spatial order of accuracy for HR and NR solvers, $t=[0,1]$ with a small time-step $\Delta t =0.001$ was used to  eliminate temporal integration errors. Computational domain $(x,y) = [0,2] \times [0,2] $ was descretized in the range of $\text{N}_x=\text{N}_y= 10, 20, 30, 40, 50$ and 60 which corresponds to $h$ as two times the circumradium of a triangle in finite element mesh, $h= 0.283, 0.141, 0.0943, 0.0707, 0.0566$. For the NR solver, the study was expanded by using meshes up to $\text{N}_x=\text{N}_y=100$. The $\mathbb{P}_1$ element was used for both velocity and pressure. As can be seen in Figure~\ref{fig_order_of_accuracy}(a), the HR and NR solvers were achieved to the second and first-order of accuracy, respectively, for both velocity and pressure.
To study temporal order of accuracy, $\mathbb{P}_4$ element and $\mathbb{P}_3$ element were used for velocity and pressure, respectively, to eliminate spatial discretization errors. $t=[0,6]$ was chosen and the following time-steps were studied, $\Delta t= 0.25, 0.125, 0.625$ and 0.03125. The study was expanded to $\Delta t=0.0078125$ for NR solver. \ref{fig_order_of_accuracy}(b) shows that the second and first-order of convergence for both velocity and pressure were obtained for HR and NR solvers, respectively.


\section{Hemodynamic indices formulations}\label{HIs_definition}

The mathematical formulations of used hemodynamic indices are presented here in Table A1, where $\tau$ is the wall shear stress vector, $T$ is end time of cardiac cycle.

 \input{Table_4}


\newpage
\section{Mesh convergence study}\label{app_grid}

\begin{figure}[!htbp] 
\centering
\includegraphics[width=0.95\textwidth]{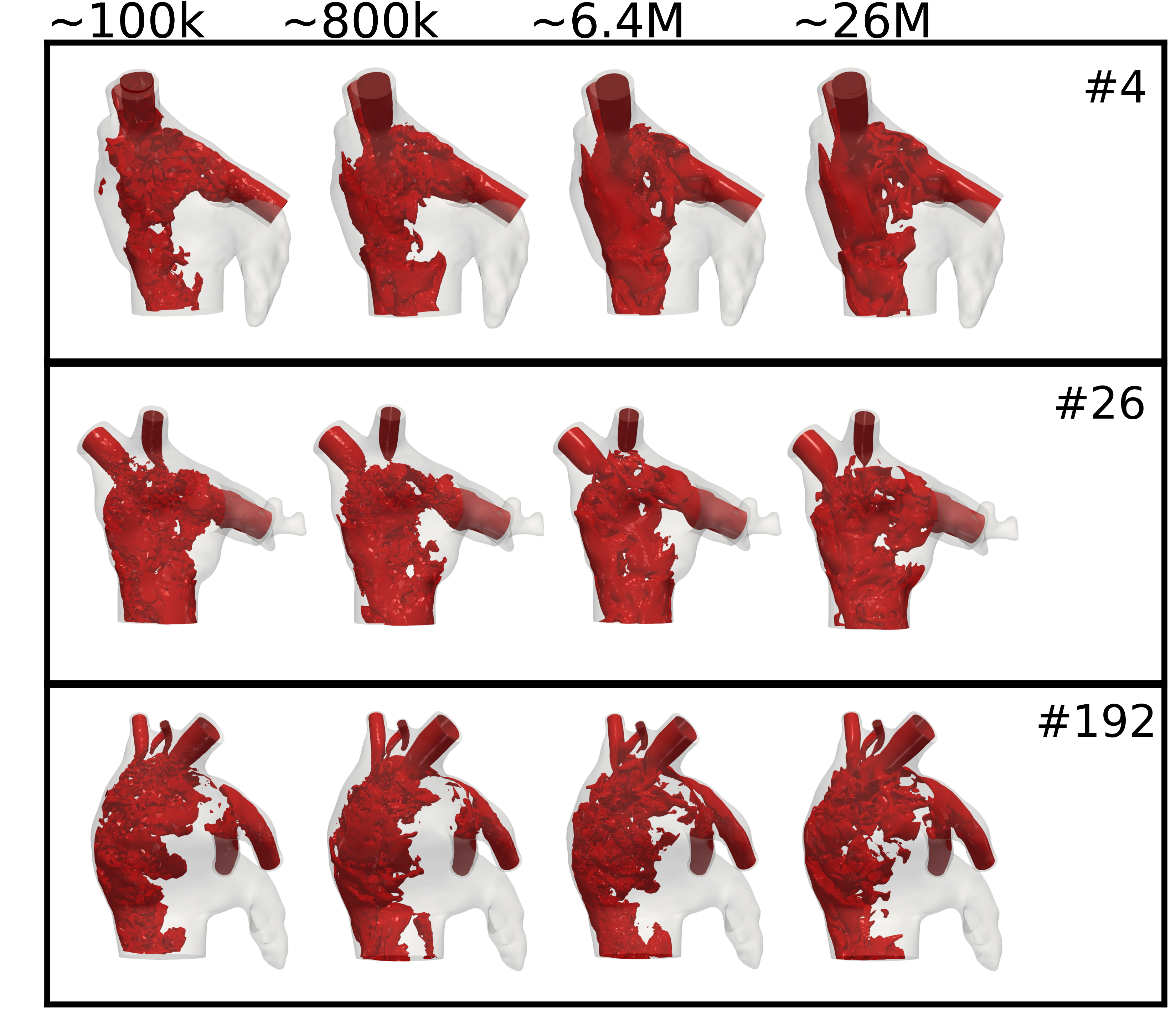}
\caption{The qualitative impact of mesh resolution (100k, 800k, 6.4M, and 26M elements) on isovelocity surfaces (in the range of [0.18-0.22] m/s) for cases 4, 26, and 192.}
\label{fig_3_isocel_mesh}
\end{figure}

\begin{figure}
\centering
\includegraphics[width=0.65\textwidth]{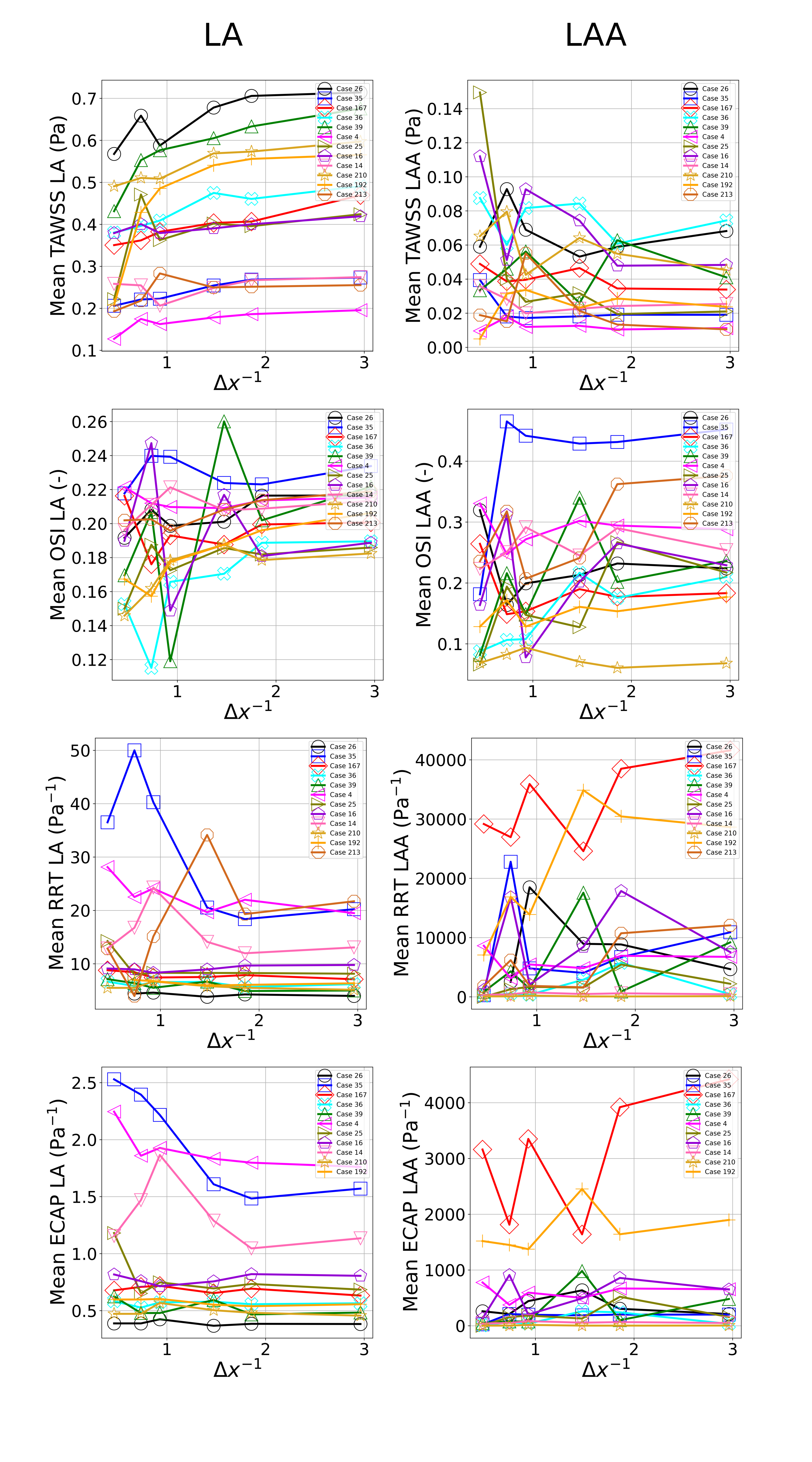}
\caption{The impact of mesh resolutions on the hemodynamic indices  time averaged wall shear stress (WSS), oscillatory shear index (OSI), relative residence time (RRT), and endothelial cell activation potential (ECAP), respectively, separated for the left atrium (LA, left) and left atrial appendage (LAA, right). $\Delta x=\sim2.4, \sim1.2, \sim0.6$, and $\sim0.4$ mm corresponds to meshes of 100k, 3.2M, 6.4, and 26M elements, respectively (cf. Table~\ref{tab_mesh_sizes}).}
\label{fig_all_results}
\end{figure}

Figure~\ref{fig_3_isocel_mesh} presents the qualitative impact of mesh resolution (100k, 800k, 6.4M, and 26M elements) on isovelocity surfaces for the cases 4, 26, and 192. 
Figure~\ref{fig_all_results} shows quantitative results of all 12 cases on all 6 meshes (i.e., 100k, 400k, 800k, 3.2M, 6.4M, and 26M elements) for WSS, OSI, RRT, and ECAP in the LA and LAA separately. Three different phenotypical behaviours in the results are identified, which classified as; (1) High variability between mesh resolutions with a staggered pattern, (2) Less variability but high errors between fine and coarse meshes, and (3) inconsiderable/low variability with a smooth pattern.

\end{document}

%% file: Table_1.tex
\begin{table}[!htbp]
  \center
  \caption{Summary of morphological characteristics of the analysed cohort. 
  }
 \begin{tabular}{l}
  \begin{adjustbox}{max width=0.85\textwidth}
  \label{tab_cohort}

    \begin{tabular}{cccccl}
    \hline
      \hline
      \\[-1em]
    Case number &  {LA}$^{1}$ volume & {LAA}$^{2}$  volume  & Number of {PVs}$^{3}$  & {LAA}$^{2}$ ostium perimeter  & {LAA}$^{2}$ morphology\\ 
   (\#)  &   (mL) &  (mL)  & (\#) &  (mm)  & \\ 
    \hline

26 &	103.23	& 3.87	& 	3 &	47.08 & Windsock \\
35 &	223.00	& 19.18 &	4 &	109.33  & Cactus \\
167	&   164.00 & 11.63 &		4 &	85.14 & Chicken wing \\
36 &	153.40 &	5.00	& 	4 &	65.68 &  Chicken wing \\
39 &	96.28 &	2.86 &	 4	& 50.16 & Cactus\\
4  &	301.03	& 26.00 &		4 &	96.97 &  Cactus \\
25 &	188.50 & 11.40 &		5 &	66.50 &  Cactus \\
16 &	182.22 &7.00	& 	5 &	90.37 &  Chicken wing\\
14 &	280.10	 & 18.99 &		5 &	120.12 &  Cauliflower\\
210 & 	113.78	& 6.50 &	 6	 &75.96 &  Cauliflower\\
192	& 222.57 &	18.17 &		6 &	113.01 &  Cactus\\
213	& 310.94 &	14.95 &		7 &	72.84 &  Cactus\\

\hline

 \end{tabular}
 \end{adjustbox}

\end{tabular}
  \begin{tablenotes}
  \footnotesize
   $^1$\ {LA: Left atrium.}\\
   $^2$\ {LAA: Left atrial appendage.}\\
   $^3$\ {PVs: Pulmonary veins.}
  \end{tablenotes}
  
\end{table}

%% file: Table_2.tex
\begin{table}[!htbp]
  \center
  \caption{Mesh characteristics. 
  }
 \begin{tabular}{l}
  \begin{adjustbox}{max width=0.9\textwidth}
  \label{tab_mesh_sizes}

    \begin{tabular}{ccccl}
    \hline
      \hline
      \\[-1em]
      Mesh size & $\Delta x$  & Nodes per volume \\ 
       (million of elements) &  (mm)  &  (1/mm$^{3}$) \\ 
    \hline

	0.1	& $\sim$2.4	& 	300-1000 \\
        0.4	& $\sim$1.8	& 	1200-4300\\
	0.8	& $\sim$1.2	& 	2500-8600 \\
	3.2	& $\sim$0.9	& 	10000-34000 \\
	6.4	& $\sim$0.6	& 	20000-68000 \\
	26	& $\sim$0.4	& 	82000-280000 \\

\hline

\ifx
    Number &  Mesh size & $\Delta x$  & Nodes per volume \\ 
    (\#) &   (million of elements) &  (mm)  &  (1/$\text{mm}^{3}$) \\ 
    \hline

1 &	0.1	& $\sim$2.4	& 	300-1000 \\
2 &	0.4	& $\sim$1.8	& 	1200-4300\\
3 &	0.8	& $\sim$1.2	& 	2500-8600 \\
4 &	3.2	& $\sim$0.9	& 	10000-34000 \\
5 &	6.4	& $\sim$0.6	& 	20000-68000 \\
6 &	26	& $\sim$0.4	& 	82000-280000 \\

\fi

 \end{tabular}
 \end{adjustbox}

\end{tabular}
\end{table}

%% file: Table_3.tex
\begin{table}[!htbp]
  \centering
  \caption{Summary of variability of modeling approaches such as spatial resolution, temporal resolution and solution accuracy. 
  }
 \begin{tabular}{c}
  \begin{adjustbox}{max width=0.99\textwidth}
  \label{tab_summary_literature}

    \begin{tabular}{lccll}
    \hline
      \hline
      \\[-1em]
     Authors, year &  Mesh sizes  & Time-steps  & Solver & Order of accuracy  \\ 
      &  (millions of elements) & per cardiac cycle  &  \\ 
    
    \hline
Zhang et al., 2008 \cite{zhang2008characterizing} &  0.17  &    -- &  In-house code &  -- \\ 

Dahl et al., 2011 \cite{dahl2012impact} &  2.2  &  2000   &  Ansys Fluent 13 & --  \\ 

Koizumi et al., 2015 \cite{koizumi2015numerical} &  0.15  &    200 &  Ansys Fluent 6.3 &  -- \\ 

Otani et al., 2016 \cite{otani2016computational} &  0.36-0.55  &   8000  &  OpenFoam, Open source   & --  \\ 

Bosi et al., 2018 \cite{bosi2018computational} &  2-3  &   1600  & Ansys CFX  &  -- \\

Masci et al.$^a$, 2018 \cite{masci2020proof} & 0.8-1.1    &   1000  &   LifeV, Open source &  VMS-SUPG stabilization, $2^{nd}$  $\mathcal{O}$ SIBE$^{*}$\\ 

Garcia--Isla et al., 2018 \cite{garcia2018sensitivity} & 0.35-0.5    &  105   &  Ansys Fluent 12.0 & --  \\ 

Dillon-Murphy et al., 2019 \cite{dillon2018modeling} &  2  & 1000     &   CHeart, In-house & SUPG stabilization, $1^{st}$  $\mathcal{O}$ BE \\ 

Jia et al., 2019 \cite{jia2019image} &  0.04-0.06  &  400   &   Ansys Fluent 18 & --  \\ 


Masci et al.$^a$, 2019 \cite{masci2019impact} & 1.8    &   2000  &   LifeV, Open source &  VMS-SUPG stabilization, $2^{nd}$  $\mathcal{O}$ {SIBE}$^{*}$\\ 

Aguado et al., 2019 \cite{aguado2019silico} &  0.2-1  &   105  &  Ansys Fluent 18.2 &  -- \\ 

Wang et al., 2020 \cite{wang2020numerical} &  2.4-5  & 100    &   Ansys workbench 16.1 &  -- \\ 

Fanni et al., 2020 \cite{fanni2020correlation} &  1.9-4.2  & 160    &   Ansys Fluent & --  \\ 

Garc{\'\i}a-Villalba et al., 2020 \cite{garcia2021demonstration} & 16.7   &  20000   & TUCAN, In-house code   &  $2^{nd}$  $\mathcal{O}$ FDM \\

Quereshi et al., 2020 \cite{qureshi2020modelling} & 0.4   &   1000  &   CHeart, In-house & SUPG stabilization, $1^{st}$  $\mathcal{O}$ BE \\

Grigoriadis et al., 2020 \cite{grigoriadis2020wall} & 3-4.9   &   100  &   Ansys CFX 15.7 & -- \\

Mill et al., 2020 \cite{mill2021silico} & 0.8-0.9   &   176  &   Ansys Fluent 19.2 & -- \\

Sanatkhani et al., 2021 \cite{sanatkhani2021subject} & 0.3-0.5   &  1600   &  OpenFoam, Open source  &  $1^{st}$ order in time, $2^{nd}$  $\mathcal{O}$  in space \\ 

D'Alessandro et al., 2021 \cite{d2020simulation} &  --    &   1000  &   LifeV, Open source &  VMS-SUPG stabilization, 2$^{nd}$  $\mathcal{O}$ SIBE$^{*}$\\ 

Fang et al., 2021 \cite{fang2021impact} & --   &   280  & Ansys Workbench 2019R3  &  -- \\ 

Mill et al., 2021 \cite{mill2021sensitivity} &  0.1-0.5  & 105    &   Ansys Fluent 19.2 & --  \\ 

Danil Vella et al., 2021 \cite{vella2021effect} & 2.8-4.5   &   --  &   Ansys CFX 19.2 &  -- \\

Paliwal et al., 2021 \cite{paliwal2021presence} &  3-8  & 920    &  OpenFoam, Open source  &   $1^{st}$  $\mathcal{O}$    \\ 

Zingaro et al.$^a$, 2021 \cite{zingaro2021hemodynamics} & 0.8-8.3   &   16000  &   LifeV, Open source &  SUPG  \\ 

Due\~{n}as-Pamplona et al., 2021 \cite{duenas2021comprehensive} &  0.8-10  &  1000   &  Ansys Fluent 2019R3 &  -- \\ 

Gonzalo et al., 2022 \cite{gonzalo2022non} &  16.7  &  20000   &  TUCAN, In-house code  &  $2^{nd}$  $\mathcal{O}$  FDM \\ 
Alinezhad et al., 2022 \cite{alinezhad2022left} &  5-7.6  &  100   &  Ansys CFX  &  --

\\

\hline
Median  value &  0.8  &  1000  &  -- &  --

\\
\hline



\end{tabular}
  \end{adjustbox}

\end{tabular}

  \begin{tablenotes}
  \footnotesize
   \item* Variational Multiscale Stabilization (VMS) Streamline Upwind Petrov Galerkin (SUPG), second-order semi-implicit backward Euler (SIBE).
   \item(--) \ Hyphen defines no information was provided in the manuscript.
  \end{tablenotes}

\end{table}

%% file: Table_4.tex
\begin{table}[!htbp]
  \centering
  \caption{Hemodynamic indices studied in the present work. 
  }
  \begin{tabular}{l}
  \begin{adjustbox}{max width=0.7\textwidth}
  \label{tab_hi}
    \begin{tabular}{lccc}
    
    \hline
      \hline
      \\[-1em]
    Hemodynamic indices &  Abbreviation & Definition & Unit \\ 
    \hline
 
Time averaged wall shear stress & WSS & $ \displaystyle \frac{1}{T} \int_{0}^{T} \lvert   \tau \lvert $dt  & (Pa) \\ 
\\
Osillatory shear index & OSI & $ \displaystyle \frac{1}{2} \left( 1 - \frac{ \lvert   \int_{0}^{T}  \tau dt \lvert }{\int_{0}^{T} \lvert   \tau \lvert dt} \right)  $  & (--) \\ 
\\
Relative residence time & RRT & $\displaystyle  \frac{1}{ (1 - 2 \cdot \textsc{OSI}) \cdot \textsc{WSS}}$  & (Pa$^{-1}$) \\ 
\\
Endothelial cell activatetion potential & ECAP & $ \displaystyle \frac{\textsc{OSI}}{\textsc{WSS}}$  & (Pa$^{-1}$) \\ 
\\
\hline
 \end{tabular}
   \end{adjustbox}


\end{tabular}
\end{table}